\documentclass[longauth]{aa}  
\usepackage{graphicx}

\usepackage{natbib,twoopt}
\usepackage[breaklinks=true]{hyperref} 
\bibpunct{(}{)}{;}{a}{}{,} 
\makeatletter
\newcommandtwoopt{\citeads}[3][][]{\href{http://adsabs.harvard.edu/abs/#3}%
{\def\hyper@linkstart##1##2{}%
\let\hyper@linkend\@empty\citealp[#1][#2]{#3}}}
\newcommandtwoopt{\citepads}[3][][]{\href{http://adsabs.harvard.edu/abs/#3}%
{\def\hyper@linkstart##1##2{}%
\let\hyper@linkend\@empty\citep[#1][#2]{#3}}}
\newcommandtwoopt{\citetads}[3][][]{\href{http://adsabs.harvard.edu/abs/#3}%
{\def\hyper@linkstart##1##2{}%
\let\hyper@linkend\@empty\citet[#1][#2]{#3}}}
\newcommandtwoopt{\citeyearads}[3][][]%
{\href{http://adsabs.harvard.edu/abs/#3}
{\def\hyper@linkstart##1##2{}%
\let\hyper@linkend\@empty\citeyear[#1][#2]{#3}}}
\makeatother

\newcommand{\teff}{\mbox{$T_{\rm eff}$}}
\newcommand{\logg}{\mbox{$\log g$}}

\newcommand{\kms}{\mbox{km\,s$^{-1}$}}

\newcommand{\degree}{\ensuremath{^\circ}}

\def\m2s2{\hbox{\,m$^{2}$\,s$^{-2}$}} 
\def\kms{\hbox{\,km\,s$^{-1}$}}       
\def\Msun{\hbox{$M_{\odot}$}}             

\def\Mjup{\hbox{$\mathrm{M}_{\rm Jup}$}}
\def\Rjup{\hbox{$\mathrm{R}_{\rm Jup}$}}

\def \t0{T$_0$}

\newcommand{\project}[1]{\textsf{#1}}

\newcommand{\emcee}{\project{emcee}}

\newcommand{\pycheops}{\project{pycheops}}
\usepackage{txfonts}
\begin{document} 

   \title{Detection of the tidal deformation of WASP-103b at $3\,\sigma$ with CHEOPS}
   \author{S.~C.~C.~Barros\inst{\ref{IA},\ref{UPorto}}\thanks{E-mail: susana.barros@astro.up.pt} 
   \and B.~Akinsanmi \inst{\ref{IA},\ref{UPorto},\ref{geneve} }
   \and G.~Boué \inst{\ref{obsparis}}
   \and A.~M.~S.~Smith \inst{\ref{dlr}}
   \and J.~Laskar \inst{\ref{obsparis}}
   \and S.~Ulmer-Moll\inst{\ref{geneve}}
   \and J.~Lillo-Box\inst{\ref{madrid}}
   \and D.~Queloz \inst{ \ref{geneve},\ref{cambridge}}
   \and A.~Collier~Cameron \inst{\ref{standrews}}
   \and S.~G.~Sousa \inst{\ref{IA}}
   \and D.~Ehrenreich\inst{\ref{geneve}}
   \and M.~J.~Hooton\inst{\ref{Bern}}
   \and G.~Bruno\inst{\ref{inaf}}
   \and B.-O.~Demory\inst{\ref{Bern}}
   \and A.~C.~M.~Correia \inst{\ref{Coimbra}}
   \and O.~D.~S.~Demangeon\inst{\ref{IA},\ref{UPorto}}
   \and T.~G.~Wilson\inst{\ref{standrews}}
   \and A.~Bonfanti\inst{\ref{graz}}
    \and S.~Hoyer\inst{\ref{lam}}
  \and Y.~Alibert\inst{\ref{Bern}}
  \and R.~Alonso\inst{\ref{iac},\ref{laguna}}
    \and G.~Anglada~Escudé \inst{\ref{ICE},\ref{IEEC}}
     \and D.~Barbato \inst{\ref{geneve}}
    \and T.~Bárczy \inst{\ref{hungary}}
    \and D.~Barrado\inst{\ref{madrid}}
  \and  W.~Baumjohann\inst{\ref{graz}}
    \and M.~Beck\inst{\ref{geneve}}
    \and T.~Beck\inst{\ref{Bern}}
  \and W.~Benz\inst{\ref{Bern},\ref{Bern2}}
  \and M.~Bergomi \inst{\ref{inaf_padova}}
    \and N.~Billot \inst{\ref{geneve}}
    \and  X.~Bonfils\inst{\ref{grenoble}}
    \and F.~Bouchy\inst{\ref{geneve}}
     \and A.~Brandeker\inst{\ref{AlbaNova}}
    \and  C.~Broeg\inst{\ref{Bern},\ref{Bern2}}
     \and J.~Cabrera \inst{\ref{dlr}}
     \and V.~Cessa\inst{\ref{Bern}}
    \and S.~Charnoz \inst{\ref{uniparis}}
    \and C.~C.~V.~Damme\inst{\ref{estec}}
    \and M.~B.~Davies \inst{\ref{lund}} 
    \and M.~Deleuil \inst{\ref{lam}} 
    \and A. Deline \inst{\ref{geneve}}
    \and L.~Delrez \inst{\ref{liege},\ref{liege2}}
    \and A.~Erikson\inst{\ref{dlr}}
    \and A.~Fortier\inst{\ref{Bern},\ref{Bern2}}
    \and L.~Fossati\inst{\ref{graz}}
    \and M.~Fridlund\inst{ \ref{leiden},\ref{onsala}} 
     \and D.~Gandolfi\inst{\ref{torino}}
    \and A.~García Muñoz  \inst{\ref{berlin}}
    \and M.~Gillon\inst{\ref{liege}}
    \and M.~Güdel\inst{\ref{vienna}}
    \and K.~G.~Isaak\inst{\ref{estec1}}
    \and K.~Heng\inst{\ref{Bern2},\ref{warwick}}
    \and L.~Kiss \inst{\ref{hungary2},\ref{hungary2},\ref{sydney}}
    \and A.~Lecavelier des Etangs\inst{\ref{instparis}}
    \and M.~Lendl\inst{\ref{geneve}}
     \and C.~Lovis \inst{\ref{geneve}}
    \and D.~Magrin\inst{\ref{inaf_padova}}
    \and V.~Nascimbeni\inst{\ref{inaf_padova}}
     \and P.~F.~L.~Maxted\inst{\ref{keele}}
     \and G.~Olofsson \inst{\ref{AlbaNova}}
    \and R.~Ottensamer\inst{\ref{vienna}}
    \and I.~Pagano\inst{\ref{inaf}}
     \and  E.~Pallé \inst{\ref{iac},\ref{laguna}}
     \and H.~Parviainen\inst{\ref{iac},\ref{laguna}}
    \and G.~Peter\inst{\ref{dlr}}
    \and G.~Piotto \inst{\ref{inaf_padova},\ref{padova}} 
    \and D.~Pollacco  \inst{\ref{warwick}}
    \and R.~Ragazzoni \inst{\ref{inaf_padova},\ref{padova}} 
    \and N.~Rando \inst{\ref{estec}}
    \and H.~Rauer \inst{\ref{dlr},\ref{berlin},\ref{berlin2}}
    \and I.~Ribas\inst{\ref{ICE},\ref{IEEC}}
     \and  N.~C.~Santos\inst{\ref{IA},\ref{UPorto}}
      \and G.~Scandariato\inst{\ref{inaf}}
    \and D.~Ségransan \inst{\ref{geneve}}
     \and  A.~E.~Simon\inst{\ref{Bern}}
    \and M.~Steller \inst{\ref{graz}}
      \and Gy.~M.~Szabó \inst{\ref{hungary45},\ref{hungary46}}
    \and N.~Thomas \inst{\ref{Bern}}
    \and S.~Udry \inst{\ref{geneve}}
    \and B.~Ulmer \inst{\ref{Frankfurt}}
    \and  V.~Van~Grootel\inst{\ref{liege}}
    \and N.~A.~Walton\inst{\ref{iacam}}
          }
 \institute{Instituto de Astrof\'isica e Ci\^encias do Espa\c{c}o, Universidade do Porto, CAUP, Rua das Estrelas, PT4150-762 Porto, Portugal \label{IA} 
            \and
            Departamento\,de\,Fisica\,e\,Astronomia,\,Faculdade\,de\,Ciencias,\,Universidade\,do\,Porto,\,Rua\,Campo\,Alegre,\,4169-007\,Porto,\,Portugal \label{UPorto}
             \and
             IMCCE, UMR8028 CNRS, Observatoire de Paris, PSL Univ., Sorbonne Univ., 77 av. Denfert-Rochereau, 75014 Paris, France \label{obsparis}
             \and
             Institute of Planetary Research, German Aerospace Center (DLR), Rutherfordstrasse 2, 12489 Berlin, Germany \label{dlr}
             \and
             Observatoire Astronomique de l'Université de Genève, Chemin Pegasi 51, Versoix, Switzerland \label{geneve}
              \and
             Depto. de Astrofisica, Centro de Astrobiologia (CSIC-INTA), ESAC campus, 28692 Villanueva de la Cañada (Madrid), Spain \label{madrid}
             \and
             Cavendish Laboratory, JJ Thomson Avenue, Cambridge CB3 0HE, UK\label{cambridge}
             \and 
             Centre for Exoplanet Science, SUPA School of Physics and Astronomy, University of St Andrews, North Haugh, St Andrews KY16 9SS, UK \label{standrews}
             \and
             Physikalisches Institut, University of Bern, Gesellsschaftstrasse 6, 3012 Bern, Switzerland \label{Bern}
             \and 
             INAF, Osservatorio Astrofisico di Catania, Via S. Sofia 78, 95123 Catania, Italy\label{inaf}
             \and
             CFisUC, Departamento de Física, Universidade de Coimbra, 3004-516 Coimbra, Portugal \label{Coimbra}
             \and
             Space Research Institute, Austrian Academy of Sciences, Schmiedlstrasse 6, A-8042 Graz, Austria \label{graz}
             \and
    Aix Marseille Univ, CNRS, CNES, LAM, 38 rue Frédéric Joliot-Curie, 13388 Marseille, France \label{lam}
     \and
    Center for Space and Habitability, Gesellsschaftstrasse 6, 3012 Bern, Switzerland \label{Bern2}
     \and
    Astrophysics Group, Keele University, Staffordshire, ST5 5BG, United Kingdom \label{keele}
    \and
    Department of Physics, University of Warwick, Gibbet Hill Road, Coventry CV4 7AL, United Kingdom \label{warwick}
     \and
    Instituto de Astrofisica de Canarias, 38200 La Laguna, Tenerife, Spain\label{iac}
     \and
     Department of Astronomy, Stockholm University, AlbaNova University Center, 10691 Stockholm, Sweden \label{AlbaNova}
     \and
    Department of Space, Earth and Environment, Onsala Space Observatory, Chalmers University of Technology, 439 92  Onsala, Sweden\label{onsala}
    \and
     Departamento de Astrofisica, Universidad de La Laguna, 38206 La Laguna, Tenerife, Spain \label{laguna}
     \and
    Institut de Ciencies de l'Espai (ICE, CSIC), Campus UAB, Can Magrans s/n, 08193 Bellaterra, Spain\label{ICE}
     \and
     Institut d'Estudis Espacials de Catalunya (IEEC), 08034 Barcelona, Spain \label{IEEC}
      \and
     Admatis, 5. Kandó Kálmán Street, 3534 Miskolc, Hungary\label{hungary}
     \and
    Université Grenoble Alpes, CNRS, IPAG, 38000 Grenoble, France \label{grenoble}
     \and 
    Université de Paris, Institut de physique du globe de Paris, CNRS, F-75005 Paris, France \label{uniparis}
     \and
     Centre for Mathematical Sciences, Lund University, Box 118, 22100 Lund, Sweden \label{lund}
     \and
     Astrobiology Research Unit, Université de Liège, Allée du 6 Août 19C, B-4000 Liège, Belgium  \label{liege}
     \and
     Space sciences, Technologies and Astrophysics Research (STAR) Institute, Université de Liège, Allée du 6 Août 19C, 4000 Liège, Belgium  \label{liege2}
     \and
    INAF, Osservatorio Astronomico di Padova,Vicolo dell’Osservatorio 5, 35122 Padova, Italy \label{inaf_padova}
     \and
    Dipartimento di Fisica e Astronomia "Galileo Galilei", Universita degli Studi di Padova, Vicolo dell'Osservatorio 3, 35122 Padova, Italy\label{padova}
     \and
     Leiden Observatory, University of Leiden, PO Box 9513, 2300 RA Leiden, The Netherlands \label{leiden}
     \and
    Dipartimento di Fisica, Universita degli Studi di Torino, via Pietro Giuria 1, I-10125, Torino, Italy \label{torino}
      \and
    University of Vienna, Department of Astrophysics, Türkenschanzstrasse 17, 1180 Vienna, Austria  \label{vienna}
      \and
     Science and Operations Department - Science Division (SCI-SC), Directorate of Science, European Space Agency (ESA), European Space Research and Technology Centre (ESTEC), Keplerlaan 1, 2201-AZ Noordwijk, The Netherlands  \label{estec1}
    \and
    Konkoly Observatory, Research Centre for Astronomy and Earth Sciences, 1121 Budapest, Konkoly Thege Miklós út 15-17, Hungary \label{hungary2}
         \and
     ELTE Eötvös Loránd University, Institute of Physics, Pázmány Péter sétány 1/A, 1117 Budapest, Hungary \label{hungary3}
        \and
    Sydney Institute for Astronomy, School of Physics A29, University of Sydney, NSW 2006, Australia\label{sydney}
    \and
    Institut d'astrophysique de Paris, UMR7095 CNRS, Université Pierre \& Marie Curie, 98bis blvd. Arago, 75014 Paris, France\label{instparis}
    \and
     ESTEC, European Space Agency, 2201AZ, Noordwijk, NL\label{estec}
    \and
    Center for Astronomy and Astrophysics, Technical University Berlin, Hardenberstrasse 36, 10623 Berlin, Germany \label{berlin}
     \and 
    Institut für Geologische Wissenschaften, Freie UniversitÃ¤t Berlin, 12249 Berlin, Germany\label{berlin2}
    \and
    ELTE Eötvös Loránd University, Gothard Astrophysical Observatory, 9700 Szombathely, Szent Imre h. u. 112, Hungary \label{hungary45}
    \and 
    MTA-ELTE Exoplanet Research Group, 9700 Szombathely, Szent Imre h. u. 112, Hungary \label{hungary46}
    \and
    Ingenieurbüro Ulmer, Im Technologiepark 1, 15236 Frankfurt/Oder, Germany  \label{Frankfurt}
    \and
    Institute of Astronomy, University of Cambridge, Madingley Road, Cambridge, CB3 0HA, United Kingdom \label{iacam}
    }

   \date{Received ??, ??; accepted ??} 
   \abstract
   { Ultra-short period planets undergo strong tidal interactions with their host star which lead to planet deformation and orbital tidal decay.}
   { WASP-103b is the exoplanet with the highest expected  deformation signature in its transit light curve and one of the shortest expected spiral-in times.  Measuring the tidal deformation of the planet would allow us to estimate the second degree fluid Love number and gain insight into the planet's internal structure. Moreover, measuring the tidal decay timescale would allow us to estimate the stellar tidal quality factor, which is key to constraining stellar physics. }    
   {We obtained 12 transit light curves of WASP-103b with the CHaracterising ExOplanet Satellite (CHEOPS) to estimate the tidal deformation and tidal decay of this extreme system. We modelled the high-precision CHEOPS transit light curves together with systematic instrumental noise using multi-dimensional Gaussian process regression informed by a set of instrumental parameters. To model the tidal deformation, we used a parametrisation model which allowed us to determine the second degree fluid Love number of the planet. We combined our light curves with previously observed transits of WASP-103b with the Hubble Space Telescope (HST) and Spitzer to increase the signal-to-noise of the light curve and better distinguish the minute signal expected from the planetary deformation.}
   {We estimate the radial Love number of WASP-103b to be $h_f = 1.59^{+0.45}_{-0.53}$. This is the first time that the tidal deformation is directly detected (at $3\, \sigma$)  from the transit light curve of an exoplanet. Combining the transit times derived from CHEOPS, HST, and Spitzer light curves with the other transit times available in the literature, we find no significant orbital period variation for WASP-103b. However, the data show a hint of an orbital period increase instead of a decrease, as is expected for tidal decay. This could be either due to a visual companion star if this star is bound, the Applegate effect, or a statistical artefact. }
   { The estimated Love number of WASP-103b is similar to Jupiter's. This will allow us to constrain the internal structure and composition of WASP-103b, which could provide clues on the inflation of hot Jupiters. Future observations with James Webb Space Telescope (JWST) can better constrain the radial Love number  of WASP-103b due to their high signal-to-noise and the smaller signature of limb darkening in the infrared. A longer time baseline is needed to constrain the tidal decay in this system.} 
   

 \maketitle
%

\section{Introduction} 
\label{intro}

The extreme environment that ultra-short orbital period planets are subjected to makes them ideal laboratories to study planetary physics. In addition to the very high temperatures, they also suffer from intense tidal forces which lead to a deformation of the planet's shape \citep{Correia2013} and shrinkage of the planet's orbit. Hence, their study allows us to gain a wealth of information on planet-to-star tidal interactions. As part of the CHaracterising ExOplanet Satellite (CHEOPS) \citep{Benz2021} Guaranteed Time Observing (GTO) programme, we are investigating the tidal interaction between ultra-hot Jupiters and their parent stars by attempting to measure their tidal decay and deformation.

Tidal forces tend to circularise planetary orbits and to synchronise the planetary and stellar rotation with the orbital period. In hot Jupiter systems, the orbits are usually circularised and the planet rotation is synchronised \citep{Ogilvie2004}. However, the synchronisation of the stellar rotation is still incomplete due to the longer and still poorly unconstrained timescale of this process. \citet{Hut1980} showed that for planets with an orbital period shorter than a third of the rotation period of the star, as it turns out to be the case for hot Jupiters, the tidal interaction leads to the unstable transfer of angular momentum from the planetary orbit to the stellar angular momentum. This results in the planet spiralling inwards and eventually being engulfed by the star. Therefore, tidal interactions between a star and a close-in exoplanet lead to shrinkage of the orbit and eventual tidal disruption of the planet. The synchronisation timescale of the stellar rotation depends on the tidal quality factor $Q'_*$ which is poorly constrained. 

The parameter $Q'_*$ allows us to constrain stellar physics \citep[e.g.][]{Ogilvie2007} and hence many attempts have been made to measure it. Studies of binary stars have estimated the tidal quality factor to be between $10^6 -10^7$ \citep{Meibom2005}. However, hot Jupiter systems could be in a different tidal regime with a higher tidal factor  ($Q'_*=10^8$) and a weaker tidal decay \citep{Ogilvie2007, Penev2011}. Estimates of $Q'_*$ through the measurement of the orbital period decrease were successful for the following hot Jupiters: WASP-4 ($Q'_*=10^4$ --\citealt{Bouma2019}) and WASP-12 ($Q'_*=10^5$ \citealt{Maciejewski2016, Yee2020}). However, the measured values of $Q'_*$ are lower than expected by theory (implying a stronger tidal dissipation) and it has not been possible to completely rule out other causes for the period decrease in these systems, such as apsidal precession. Statistical studies of the ensemble of known hot Jupiters show two regimes of tidal dissipation strength. The majority of the studied systems had $\log_{10} Q'_* = 8.26 \pm 0.14$, while a smaller group had $\log_{10} Q'_* = 7.3\pm 0.4$ \citep{CollierCameron2018}.

The tidal deformation of a planet mostly depends on the planet-to-star distance and it is most significant for large planets that are almost filling their Roche lobe \citep[e.g.][]{Ferraz-Mello2008}. Hence, it is larger for ultra-hot Jupiters. The radial deformation of a planet due to a perturbing potential can be quantified using the second degree fluid Love number $h_f$ \citep{Love1911}. The Love number measures the distribution of mass within the planet depending on the concentration of heavy elements in the core of the planet relative to the envelope of the planet. Therefore, it provides insight into the internal structure differentiation of planets \citep{Kramm2011}. \citet{Correia2014} shows that $h_f$ is proportional to an asymmetry parameter $q$ which relates the three axes ($r_1$, $r_2$, and $r_3$) of the planetary ellipsoidal shape -- if $r_1=r_2(1+3q) $ and $r_3=r_2(1-q) $, then  
\begin{equation} 
\label{hf}
h_f =  2q \frac{M_p}{M_{\star}} \left(\frac{a}{R_*}\frac{1}{R_V} \right)^3, 
\end{equation} 
with the volumetric radius $R_V =\sqrt[3]{r_1r_2r_3}$ $M_p$ and $M_{\star}$ being the planetary and the stellar mass, respectively, $R_*$ being the stellar radius, and $a$ being the semi-major axis of the planet's orbit.   \citet{Correia2014} also shows that the non-spherical shape of a deformed planet along with its varying projected area during a transit modifies the transit light curve and causes anomalies in the ingress, egress, and mid-transit phases compared to a spherical case (Figure~A1 of \citealt{Correia2014}). Detecting the deformation-induced signature in the light curve can therefore allow for the measurement of the Love number \citep{Akinsanmi2019, Hellard2019}.

Of the several attempts made to measure the deformation signature, the most constraining is for WASP-121 using two Hubble Space Telescope (HST) /Space Telescope Imaging Spectrograph (STIS) transits ( $h_f = 1.39\pm 0.8$ -- $< 2  \sigma$ significance -- \citealt{Hellard2019}). A measurement of an exoplanet's Love number was made for HAT-P-13b \citep{Buhler2016}. HAT-P-13b has a unique orbital configuration that allows for the measurement of the Love number using apsidal precession. However, in this case some assumptions are required to estimate the Love number (see Section~\ref{precession}). \citet{Batygin2009} constrained the Love number of HAT-P-13b to be $1.116 < h_f < 1.425$. This was later updated to $h_f =1.31^{+0.08}_{-0.05} $ and allowed constraints on the maximum core size and the metallicity of the planet’s envelope, showing the power of the Love number in providing insights into the internal structure of the planet \citep{Kramm2012, Buhler2016}. Furthermore, unveiling the internal structure is in turn important for understanding the formation of the planet itself since the distribution of the heavy elements and the core mass directly depend on formation mechanisms \citep{Mordasini2012}. 


Taking advantage of the high-precision and high pointing flexibility of the CHEOPS satellite, we designed a programme to measure the tidal decay and deformation of ultra-hot Jupiters. The expected amplitude of the deformation signature is largest for WASP-103b  ($\sim 60$ ppm) due to its larger radius among the ultra-hot Jupiters. Hence, this target was a priority for our programme. WASP-103b is a 1.5 \Mjup\ and 1.5 \Rjup\ planet in a 22 hour orbit around a late F-type star with a G magnitude of 12.2 \citep{Gillon2014}. 
 The small amplitude of the tidal deformation signal has prevented its detection until now and requires that CHEOPS transits are combined with other high signal-to-noise transits in order to allow us to estimate the planet's Love number. The required long baseline of observations to measure the tidal decay of exoplanets also requires that the derived transit times from CHEOPS are combined with previously derived transit times.

In this paper, we present the first results of our tidal decay and deformation programme targeting WASP-103b. In Section~2 we describe the CHEOPS observations and in Section~3 we describe complementary observations necessary to better constrain the system. In Section~4, we present our results for the variation of the planetary orbital period and discuss possible scenarios to explain it. In Section~5, we present our modelling of the tidal deformation combining CHEOPS results with HST and Spitzer observations. Finally, we draw our conclusions in Section~6.

\section{Observations, data reduction, and analysis}

\subsection{CHEOPS observations of WASP-103b}

The objective of CHEOPS is to achieve a detailed characterisation of known exoplanets through high-precision photometric observations.
It is the first S-class ESA mission and it was launched on 18\ December 2019 \citep{Benz2021} with science observations starting in April 2020.
We obtained data as part of the CHEOPS Guaranteed Time Observing (GTO) programme: 'Tidal decay and deformation (ID 0013)'. This programme aims
to measure the tidal deformation and decay of short period exoplanets in order to constrain the planetary Love number and the stellar tidal dissipation parameter. This programme is included in one of the six GTO themes called feature characterisation which also includes one programme to search for moons and rings and one programme to measure the angle between the planetary orbit and the stellar spin through the gravity darkening effect.

Currently the tidal deformation programme includes the targets WASP-12b and WASP-103b. These, together with WASP-121b, are the best known targets to measure the tidal deformation directly from the light curve \citep{Akinsanmi2019}. Unfortunately, WASP-121b is not observable by CHEOPS due to pointing restrictions.

Due to the extremely high photometric precision necessary to measure the tidal deformation, the original plan was to obtain 20 transits per year over the 3 years of the GTO. Due to the best visibility and observational efficiency of WASP-103 compared with WASP-12, this target was given priority and 20 transits were requested in the first year of the CHEOPS nominal mission. CHEOPS data suffer from interruptions due to Earth occultations or passages through the South Atlantic Anomaly (SAA), which can affect our observations. Therefore, extensive tests were performed during the preparation of the GTO of CHEOPS that show that a good coverage of ingress and egress is crucial in order to obtain accurate and precise times and also to better sample the shape deformation. 

We requested 90\% efficiency in ingress and egress and 60\% overall efficiency of transit observations. Since this would decrease the number of possible observable transits, we started by requesting 90\% efficiency in either ingress or egress. However, the first three transits showed poorer precision of the derived transit times, and hence, we included a stronger constraint of having 90\% efficiency in both ingress and egress. This allowed us to observe only 12 of the 20 requested transits, but with increased accuracy for the derived transit times. After the first three test observations, we also increased the total requested time per observation. Originally, we requested the observations to cover three transit durations. For WASP-103, this corresponds to $\sim 3.4$ CHEOPS orbits ($\sim$ 7.8 hours). However, for observations with an efficiency of less than 88\%,  we do not have the recommended three CHEOPS orbits of data to be able to detrend the systematic noise \citep{Maxted2021}. Hence, we increased the duration of the observations which resulted in a much better detrending of the systematic noise (see Section~\ref{datareduction}). The observation log is presented in Table~\ref{table:observations}.

\begin{table*} 
\centering 
\caption{Log of CHEOPS observations of WASP-103b. \label{table:observations}} 
\centering 
\begin{tabular}{ c c  c c  c c c c} 
\hline 
\hline 
\# &  Start date  & Duration  & $N_{obs}$ & Effic. & APER &$R_{ap}$ &Decorrelation\\ 
& (UTC) &  (hours)  &   & (\%)&  & (pixels) \\ 
\hline 
1 & 2020-04-18T22:55:40.965587 & 6.02 & 269 & 74 & OPTIMAL &17.0 & roll , bg \\   
2 & 2020-05-02T19:53:40.996584 & 6.14 & 296 & 80& DEFAULT &25 & roll  \\    
3 & 2020-05-05T14:10:00.434364 & 6.55 & 308 & 78& OPTIMAL &17.5 & bg \\  
4 & 2020-05-16T14:57:01.026698 & 9.64& 523 & 90& DEFAULT  & 25 & roll, x , y \\                
5& 2020-05-19T10:08:00.408947 & 9.67 & 544 & 94&  OPTIMAL&19.0 &roll , bg \\          
6 & 2020-05-25T21:19:00.404138 & 9.97 & 560 & 94 & OPTIMAL &19.0 & roll, x \\             
7 & 2020-06-06T22:07:00.904021 & 9.37 & 540 &96 & DEFAULT  & 25.0 & roll  \\     
8 & 2020-06-07T20:04:00.500261 & 9.64 & 546 & 94 & DEFAULT &25.0  & roll \\        
9 & 2020-06-14T08:08:00.511905 & 9.64& 533 & 92 & OPTIMAL &19.0 & roll , bg\\    
10 & 2020-06-18T23:06:00.996556 & 9.64 & 555 & 96 &  DEFAULT &25.0 &roll   \\
11& 2020-06-19T21:02:00.713493 & 9.64 & 538 & 93 &OPTIMAL &19.0 & roll , bg \\             
12& 2020-06-20T19:07:39.419756 & 9.55 & 537 & 93 & OPTIMAL& 18.5 &roll , cont,  x, bg\ \\
\hline 
\hline 
\end{tabular} 
\tablefoot{\\
The transits are labelled by their sequence number throughout the paper. Effic. is the proportion of the time in which unobstructed observations of the target occurred. $R_{ap}$ is the aperture radius used for the photometric extraction. We also give the decorrelation parameters used for each light curve roll angle (roll), background (bg), x centroid (x), y centroid (y), and contamination (cont).}
\end{table*}

\subsection{Photometric extraction}
\label{datareduction}
\begin{figure} 
\centering 
\includegraphics[width=0.9\columnwidth]{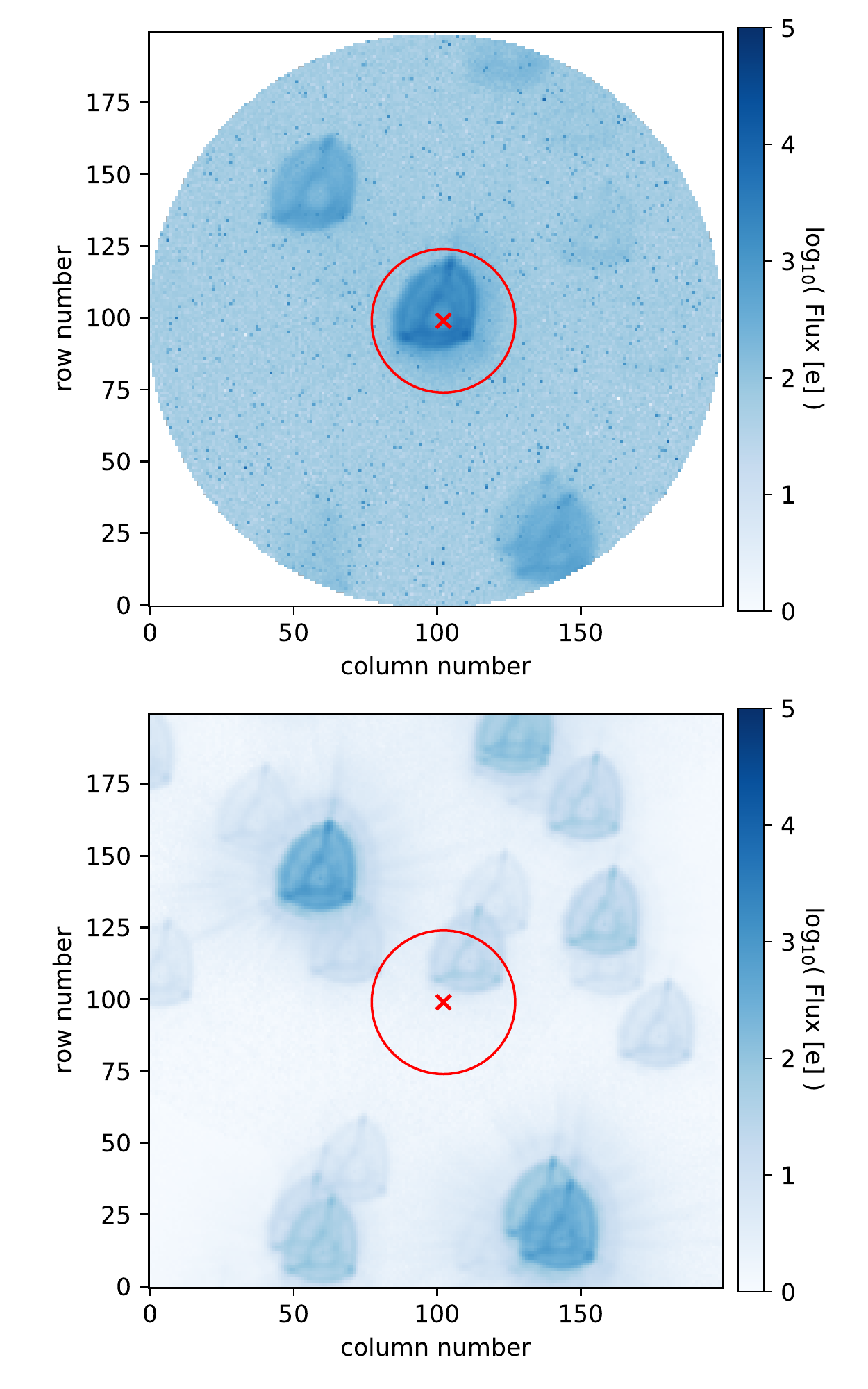}
\caption{ Top panel: Field of view of WASP-103 as observed by CHEOPS transit \#10. Bottom panel: Field of view of WASP-103 simulated by the DRP based on the \textit{Gaia} catalogue and excluding WASP-103. The red cross marks the target position, while the red circle shows the DEFAULT aperture. The image scale is 1 arcsec per pixel. \label{field}} 
\end{figure}

The CHEOPS observations were reduced with the CHEOPS data reduction pipeline (DRP) \citep{Hoyer2020}. The DRP automatically processes all the CHEOPS data. It makes bias, dark, and flat corrections, and it applies gain, scattered light, and a correction for the non-linearity of the detector response. The DRP simulates the field of view using the magnitudes and positions of stars in the \textit{Gaia} DR2 catalogue \citep{gaiadr2}. These simulations are used to calculate the contamination of the target aperture by nearby stars. Due to the irregular PSF shape coupled with the rotation of the field of view of CHEOPS, the target star suffers from variable contamination from nearby stars. This contamination is a function of the angle of rotation of the satellite (roll angle) and the pointing jitter. The DRP calculates and provides the contamination of the target aperture as a function of time so it can be corrected later. In the case of WASP-103, the simulation of the field of view shows a contaminating star inside the aperture $\sim16$ arcsec from the target, as is shown in Figure~\ref{field}. This contaminant adds $ \sim 0.9$ \% to the total flux in the aperture.

The DRP also corrects the smearing trails of bright stars in the field of view.  Due to the rotation of the field of view, this leads to a variable contamination of the target aperture.
 The DRP extracts the photometry for four apertures, three with a fixed radius (22.5, 25, and 30 pixels) and one with an optimal radius, labelled RINF, DEFAULT, RSUP, and OPTIMAL. The radius of the optimal aperture is calculated for each data set to maximise the signal-to-noise ratio of the light curve. The DRP also corrects the background light which is estimated from an annulus around the target. The DRP produces a fits file with four extracted light curves together with auxiliary information including, for example, the time series of the roll angle, the estimated contamination, the subtracted background, a quality flag,  and the centroid position of the target star. These can be used to correct any systematic effects in the light curves. Furthermore, the DRP produces a report that states the performance of each step of the pipeline. More information about the data reduction pipeline can be found in \citet{Hoyer2020}. For WASP-103, we considered both the OPTIMAL and DEFAULT apertures and chose the one with lower residuals in the final analysis as explained in the next sub-section.

\subsection{CHEOPS data analysis}

\begin{figure} 
\centering 
\includegraphics[width=0.9\columnwidth]{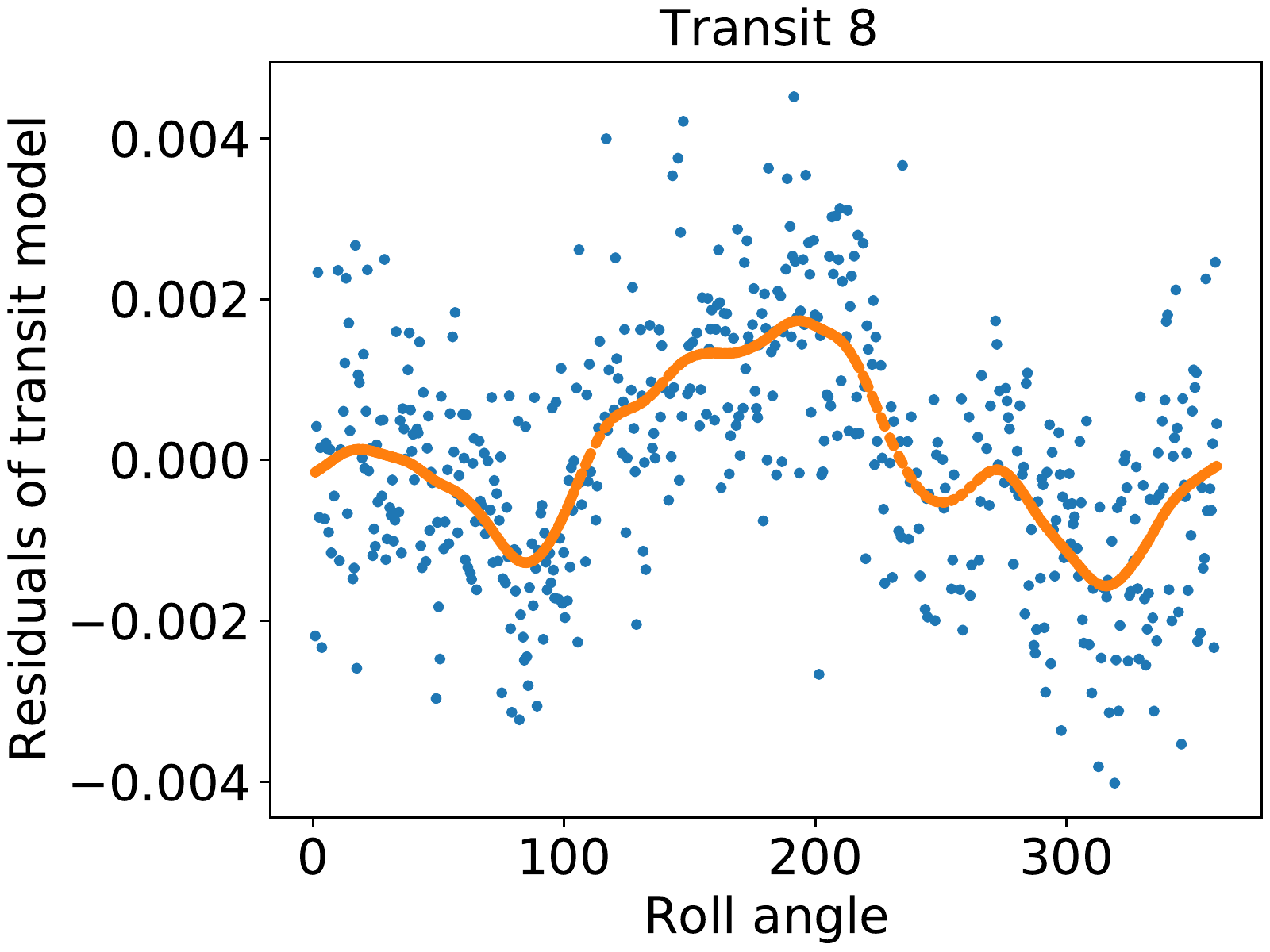}
\caption{Residuals of the transit fit as a function of the roll angle of the satellite for transit number 8. A clear dependence is seen which is well fitted with a non-parametric GP model overplotted in orange (see Section~\ref{dataanali}). The best fit length scale hyper-parameter in this case was $73^{+26}_{-10} $ degrees. The flux dependence with a roll angle is clearly seen in all our light curves, but the shape of the dependence varies.  \label{rollangledep}} 
\end{figure}

The light curves obtained with the DRP were corrected for the estimated contamination of the aperture in each light curve. Some of our observations were also affected by the atmospheric air glow at the beginning and end of an Earth occultation. In these cases, the air glow contaminates the observation, increasing the background to values higher than the target and ultimately leading to saturation.  These points cannot be corrected for and we removed all of those with a background noise higher than the median noise of the target star. We also removed outliers using $5\, \sigma$ clipping. In our case, 
all the light curves show a strong correlation with the roll angle (Figure~\ref{rollangledep}). Moreover, the light curves show extra correlations with a mix of instrumental parameters, for example, the background flux, contamination rate, and x position. These light curves are presented in Figure~\ref{LCplots}. They clearly show systematic noise which is the residual of the variable contamination of the aperture, mentioned above, and it is highly correlated with instrumental parameters such as the roll angle of the satellite, the position of the star, and the background flux. 

During the preparation of the CHEOPS mission, several methods were tested to correct the systematic noise due to the rotating field of view and it was concluded that a better accuracy is achieved if the systematics are corrected simultaneously with the transit modelling. In order to derive transit parameters and simultaneously decorrelate the CHEOPS light curves, we used two methods. The first one is based on multi-dimensional Gaussian processes (GPs) \citep{Rasmussen2006} coupled with the batman transit model \citep{Kreidberg2015}. The second method is based on linear decorrelation using a combination of sinusoidal functions implemented in \pycheops\ \citep{Maxted2021}.

\subsubsection{Multi-dimentional GP}

\label{dataanali}
We performed GP regression with a Matern-$3/2$ kernel to model the flux dependence on the instrumental parameters using the \project{George} package \citep{Ambikasaran2015}. This is coupled with a transit model using a parametrisation and method similar to the one used in \citet{Barros2020}. The transit model is parameterised by the orbital period ($P$), mid-transit time ($T_{0}$), normalised separation of the planet ($a/R_{\star}$), the planet-to-star radius ratio ($r_{p}/R_{\star}$), inclination ($inc$), and quadratic limb darkening law.  We assumed a circular orbit (\citealt{Gillon2014,Delrez2018} and Section~\ref{RV}). The hyper-parameters of the GP are an amplitude ($amp$) and a length scale ($s$).

For the shape parameters of the transit ($a/R_{\star}$, $r_{p}/R_{\star}$, and $inc$), we used Gaussian priors based on the results of \citet{Delrez2018}. For the limb darkening parameters, we also used Gaussian priors whose values and the uncertainties were derived with the LDTK code \citep{Parviainen2015,Husser2013} in the CHEOPS bandpass.
For the mid-transit time, we used a uniform prior and we assumed the ephemerides derived by \citet{Patra2020} ($T_{0} =2456836.29630(07)$ and $P = 0.925545352(94)$). For the hyper-parameter length scale, we used a uniform prior based on the range of the instrumental parameter variations. For each instrumental parameter, we computed the maximum range of variations and set this as the maximum possible value of the prior and the minimum was set to be one-sixth of this value to avoid over-fitting. For example, in the case of the roll angle, the prior is uniform between 60\degree and 360\degree. We also included a jitter parameter for each light curve. The derived values for the jitter indicate that the errors provided by the DRP pipeline are slightly underestimated by a factor of approximately 1.2.

For each transit, we started by using a 2D GP with the roll angle as the detrending instrumental parameter, then we analysed the correlations of the residuals with the other instrumental parameters and added the instrumental parameter with the highest correlations to a 3D GP. We performed model comparison to decide if the instrumental parameter should be added or not. Instrumental parameters were added if the difference of BIC was higher than 3, indicating positive evidence in favour of the more complex model. The procedure was repeated until the residuals showed correlations with the instrumental parameters less than 5\% or they were already tested with model comparison. The first three light curves were highly correlated with the background and hence we tested a 2D GP with only the background as a detrending instrumental parameter. This was found to be better only for light curve number 3. This procedure was applied to both the OPTIMAL and the DEFAULT apertures and we chose the one with smaller final residuals. The instrumental parameters chosen with this procedure for the decorrelation of each of the transit light curves as well as the aperture chosen are given in Table~\ref{table:observations}.

For parameter inference, we used the affine-invariant Markov-chain Monte-Carlo ensemble sampler implemented in \emcee\ \citep{Goodman2010,Foreman-Mackey2013}. The fitting procedure was performed in two steps. First we performed a global fit using previously normalised light curves (first order polynomial based on out-of-transit data) and assuming a linear ephemerides and the best detrending GP model. From the global fit, we derived the transit parameters and their uncertainties similarly to \citet{Barros2020}. These are given in Table~\ref{prior_transit} together with the priors used. In the second step, the posterior of the shape parameters $a/R_{\star}$, $r_{p}/R_{\star}$ and $inc$ derived from the first step were used as priors for a second individual fit to each light curve to derive accurate transit times. In this second step, we simultaneously accounted for a linear normalisation of the transit parameterised by an out-of-transit level ($F_{out}$) and flux gradient ($F_{grad}$). Fitting the transit normalisation is important to avoid biases in the derived mid-transit times as shown in \citet{Barros2013}. For each light curve, we used the best model GP determined in the previous step with the same hyper-parameters'  priors mentioned above. The two step approach was adopted because the correlations are different for each light curve and the detrending instrumental parameters considered are also different. Therefore, a simultaneous fit of the detrending instrumental parameters would imply a prohibitive number of parameters to fit (317). We performed tests that showed that this approach does not affect the results. 

\begin{figure*} 
\centering 
\includegraphics[width=2.0\columnwidth]{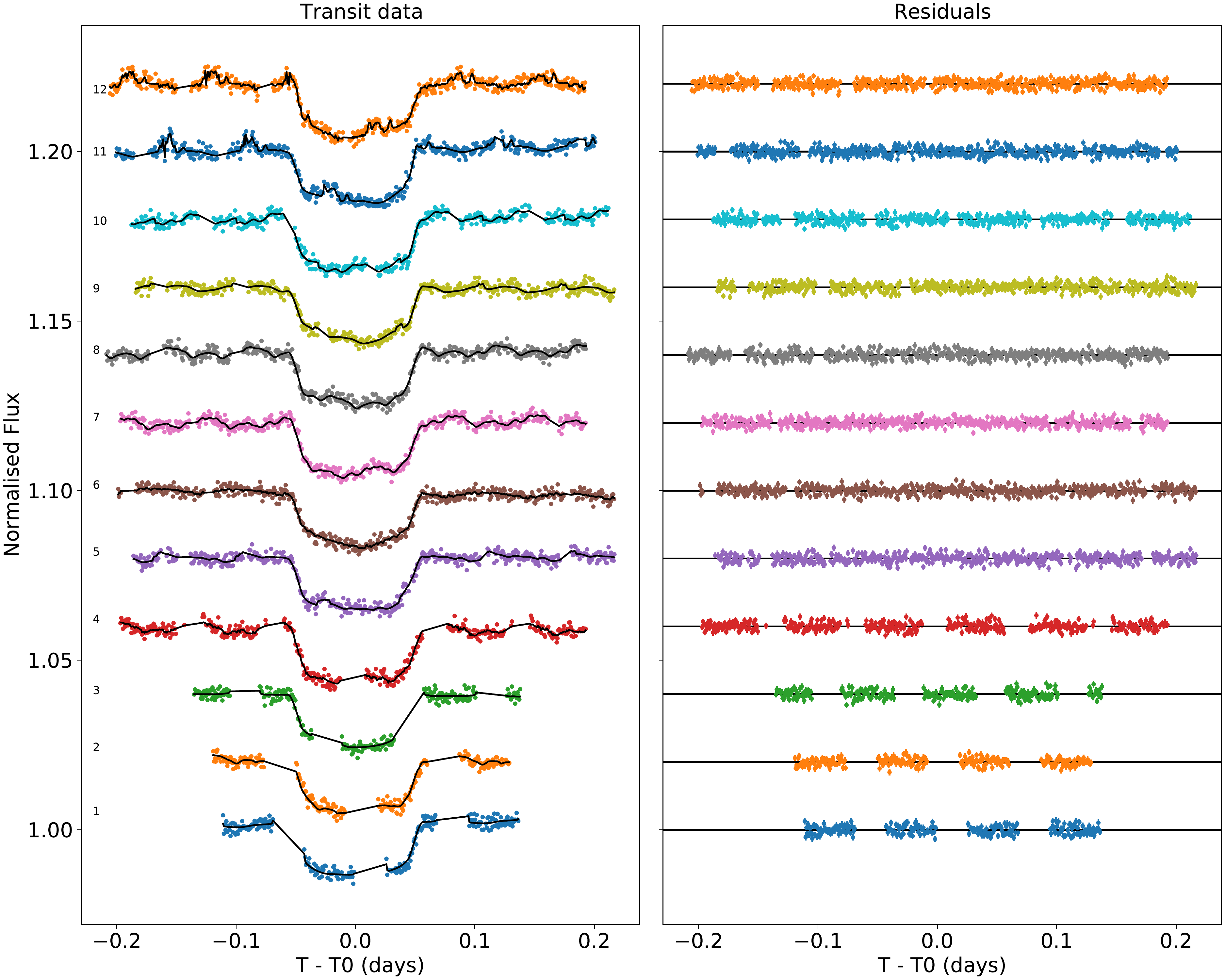}
\caption{Left panel: Transit light curves of WASP-103b obtained with CHEOPS. We overplot our best fit model that includes a transit model and the GP model to account for systematics dependent on the instrumental parameters. Right panel: Residuals of the fit of the transit model and the GP model. For clarity the errors are only shown in the residuals. The light curves and residuals are offset vertically for clarity. \label{LCplots}} 
\end{figure*}

The WASP-103b transit light curves together with the best multi-dimensional GP and transit model, chosen by model comparison, are shown in Figure~\ref{LCplots}. The light curves show instrumental effects that are well corrected by the GP model, as can be seen from the well behaved residuals.

\begin{table} 
\caption{Priors for the fitted transit parameters. \label{prior_transit} }
\begin{tabular}{lcc} 
\hline 
\hline Parameter & Prior & Derived value \\ 
\hline T$_{0}- \mathrm{T_{ref}}$ (days)   &  $\mathcal{U}(-0.1;0.1)$   & $0.000213 \pm  0.000062$ \\ 
$R_p/R_{\star}$  & $\mathcal{N}( 0.1150,0.0020)$  & $0.11100\pm 0.00014$ \\ 
$a/R_{\star}$  &  $\mathcal{N}( 3.010,0.013)$    &  $2.9829\pm 0.0054$\\ 
$inc$ [\degree]  &  $\mathcal{N}( 88.8,1.1)$  & $89.22\pm 0.60$ \\ 
$LD1$ &  $\mathcal{N}( 0.5269, 0.0218)$  & $0.5269 \pm 0.0218$ \\ 
$LD2$ &  $\mathcal{N}(  0.1279, 0.046)$  & $0.1279 \pm 0.046$ \\
\hline 
\hline 
\end{tabular} 
\tablefoot{\\
$T_{ref} =2456836.2963007 $, $\mathcal{U}(a;b)$ is a uniform distribution between $a$ and $b$;  $\mathcal{N}(a;b)$ is a
normal distribution with mean $a$ and standard deviation $b$. }
\end{table}

\subsubsection{Pycheops}
For comparison, we also analysed the OPTIMAL extracted light curves with \pycheops\  \citep{Maxted2021}, a custom python package developed specially for CHEOPS data. First we used the single visit analysis to determine the best parameters to use as decorrelation instrumental parameters.  \pycheops\ performs linear decorrelation with several instrumental parameters such as contamination, background, position of the target on the CCD, and trigonometric functions of the roll angle and its harmonics. As in the multi-GP method, we used priors based on the transit parameters derived in \citet{Delrez2018}. 

After analysing the data with the single visit model, we used the multivisit mode of \pycheops\ to simultaneously fit the 12 transits and determine the individual transit times.  To avoid the large number of fitted parameters, \pycheops\  has implemented a technique \citep{Luger2017} to perform an implicit decorrelation of several light curves using a GP. A detailed description of  \pycheops\ and example applications to CHEOPS data are given in  \citet{Maxted2021}. The derived transit times with \pycheops\ are closer than  $1\,\sigma$ to the ones derived with the multi-dimensional GP. The derived uncertainties in the transit times are also similar.

\section{Complementary observations}

\subsection{Constraining the companion of WASP-103}

\label{companion}

Using the lucky imaging technique, a possible companion to WASP-103 was detected at $0.242 \pm 0.016$ arcsec by \citet{Wollert&Brandner2015} on 07 March 2015. They measured the position angle to be  $132.7 \pm 2.7$  degrees and contrast magnitudes to be  $\Delta i = 3.11 \pm 0.46$  and  $ \Delta z  =2.59 \pm 0.35$. These observations were made with the AstraLux instrument \citep{Hormuth2008}. This companion was later confirmed by adaptive optics (AO) observations using the NIRC2 instrument at Keck \citep{Ngo2016} on the 25 January 2016. They clearly detect a companion at  $0.239 \pm 0.002$ arcsec. Within the errors, no change of position was detected between the two observations.  These observations were used by \citet{Cartier2017} to perform a spectral energy distribution
(SED) fit to the companion and estimate its parameters. This assumes that the companion is located at the same distance as WASP-103 which has not been confirmed. The stellar parameters derived by \citet{Cartier2017} together with the position measurements derived by \citet{Ngo2016} are reproduced in Table~\ref{parcompanion}. According to the AO measurements, if the companion is at the same distance as WASP-103A  ($552 \pm 33$  parsec -- \textit{Gaia} EDR3 \citep{Gaia2021} - Table~\ref{stellarp} ), it would be at $131.9 \pm 8$au from WASP-103. If the orbit is circular, this would imply a period of $1114\pm  103$ years and a radial velocity (RV) signature with an amplitude of $ \sim 1334 $ m/s.

There is an excess of astrometric noise in the \textit{Gaia} data that is consistent with the existence of a companion for this system. This noise was present in the DR2 data release and in the recent EDR3. Furthermore, the \textit{Gaia} derived parallax changed from $1.14\pm 0.17$ in DR2 to $1.82 \pm 0.11$ in EDR3 which is a $3.4\, \sigma\ $ change, which could be due to a deviation from single-source behaviour induced by a companion. Therefore, the companion still seems to be close to WASP-103 at the present epoch.

\begin{table}[ht!]
\caption{Parameters for the companion of WASP-103A.  \label{parcompanion} }
\begin{center}
\begin{tabular}{l  c }
\hline
\hline
Parameter & Value and uncertainty\\
\hline
Effective temperature \teff\ [K]  & 4330  $\pm$ 100\\
Surface gravity \logg\ [g cm$^{-2}$]  &  4.604 $\pm$  0.016\\
Spectral type & K5 V\\
Stellar mass $M_{\star}$ [M$_\odot$] &0.721 $\pm$ 0.024 \\
Stellar radius relative to A $R_{B}/R_{A}$ & 0.52  $\pm$0.05\\
Distance to WASP-103 A [mas]  & $240.0 \pm 1.5$ \\
Distance to WASP-103 A [au]  & $131.9 \pm 8 $ \\
Position angle PA  [degrees] &  $131. 37 \pm 0. 35 $\\
\hline
\hline
\end{tabular}
\tablefoot{\\
The stellar parameters of the companion of WASP-103 were derived from a SED fit assuming the same distance as WASP-103A  \citep{Cartier2017}. The position of the companion relative to WASP-103A was derived by \citet{Ngo2016} . 
}
\end{center}
\end{table}

\subsubsection{Lucky imaging observations}

To better constrain the companion of WASP-103 and confirm that it is bound, we performed new lucky imaging observations of WASP-103.
We used the same instrument that discovered the companion \citep{Wollert&Brandner2015}, the  AstraLux camera \citep{Hormuth2008} mounted on the 2.2-metre telescope at Calar Alto observatory in Almer\'{i}a, Spain. The observations were performed under excellent conditions (seeing $0.7$ arcsec) in two filters SDSSi and SDSSz. We obtained 90,000 frames, each with an exposure time of 10\,ms.

As shown in Figure~\ref{HRimage} and \ref{Sensitivity} of the appendix, we did not recover the companion found by \cite{Wollert&Brandner2015} and subsequently characterised \citet{Ngo2016} and \citet{Cartier2017}. We computed the sensitivity limits for our images by using the approach described in \cite{Lillo-Box2012,Lillo-Box2014}. According to our contrast curve, we should have detected the companion if it was in the same location (separations and position angle). No difference in the position of the companion was detected between the two previous observations  \citep{Wollert&Brandner2015,Ngo2016}, but they were only separated by 10 months.  Assuming that the target observed by us and the previous publications is the same, in order to explain the non-detection of the companion in our AstraLux images, the companion should have moved towards WASP-103A by at least $\sim 0.14$ arcsecs, which corresponds to $\sim77$ au at the distance to WASP-103 (Table~\ref{stellarp}). 

This is difficult to explain in a scenario where the companion is bound to WASP-103 since the time difference between the observations is just 6 years. The maximum velocity of a bound object is lower than the velocity at the periapsis given by the following: 
\begin{equation} 
\label{vmax}
v^2_{max} = \frac{  2G(M_{*}+M_{comp} ) }  {d}
\end{equation} 
\citep{Murray1999}, where $G$ is the gravitational constant and $d$ is the distance of the periapsis. Assuming the periapsis is equal to $131.9$au (Table~\ref{parcompanion}), in 6 years, the upper limit on the distance travelled by a bound object is $6.4 \pm 0.2$ au. Therefore, we conclude that either the star is not bound (and hence we are seeing the relative proper motion of both stars, with the background star disappearing behind WASP-103) or much less likely, unknown systematics have prevented its detection in the new observations. Further high resolution images of WASP-103 or the \textit{Gaia} DR3 will shed light on this system.

\subsubsection{CORALIE RV observations}
\label{RV}
If WASP-103A has a bound stellar companion, we expect a long-term variation in the observed RVs.  Previous RV observations of WASP-103 from the discovery paper and a follow-up paper \citep{Gillon2014,Delrez2018} do not show any long-term variation, but they only span 450 days. To constrain the longer-period RV variations, we obtained new RVs of WASP-103 with CORALIE, the same instrument that was used in previous studies. Ten new observations were made between 18 March 2021 and 30 May 2021 which increased the time span of the observations to 8 years.

The CORALIE spectrograph is installed at the Nasmyth focus of the Swiss Euler 1.2m telescope \citep{Queloz2000} and has a spectral resolution of 60,000. The light can be injected through two fibres allowing it to observe the science target and to perform a simultaneous monitoring of the sky or a wavelength calibration with a Fabry Perot etalon. In November 2014, the spectrograph benefited from a major upgrade, which introduced an RV offset which we modelled as a simple offset between the two sets of data. The observations were processed with the standard data reduction pipeline. The RVs were derived with the weighted cross-correlation technique \citep{Pepe2002} and a G2 mask was used as it is optimised for late F-type stars such as WASP-103. 

The RVs were analysed with the code LISA \citep{Demangeon2018,Demangeon2021} which uses the \project{radvel} python package \citep{Fulton2018} to model the RV observations. The RV model is parameterised by the semi-amplitude of the RV signal (K), the planetary period (P), the mid-transit time ($T_0$), and the products of the planetary eccentricity by the
cosine and sine of the stellar argument of periastron $e\cos{\omega}$, $e\sin{\omega}$. We used Gaussian priors for the planetary orbital period and mid-transit time based on the constant period model derived in Section~\ref{ephemerides}. We included an offset between the new RV observations and the previously published RV observations to account for the RV shift due to the upgrade of the instrument mentioned above. We also included a jitter parameter for each dataset to account for unknown systematic noise or short-term stellar activity. We compared a model with a beta prior on the orbital eccentricity \citep{Kipping2013} with a circular model. Due to the short time span of the observations compared with the expected orbital period of the possible binary, the visual companion, if bound, would lead to a trend in the RV observations. Moreover, given the short span of the two epochs of observations and the large gap between them, this trend can be represented by an offset between both observations. Therefore, the fitted offset is a combination of the instrumental offset and the trend due to the possible companion star.  

With the free eccentricity model, we found a non-significant eccentricity of $0.11 \pm 0.06$. The difference between the Bayesian information criterion (BIC) of the eccentric and circular model is 0.011 which implies that the eccentric model is not justified. As mentioned above, due to the very short orbital period of WASP-103b, the orbit of the planet is expected to be circularised and the rotation of the planet synchronised with the orbital period. Therefore, we adopted the circular model for the planet. We found a semi-amplitude $K = 268 \pm 14 $ m/s in agreement with previous results \citep{Delrez2018}. We also found an offset between the previous observations and the new observations of $14 \pm 45$ m/s.  At $3\, \sigma $ we can exclude an offset higher than $151$ m/s and lower than  $-119$ m/s. The relative offset due to the instrumental upgrade between the two observations is expected to be between 14 and 24 m/s (CORALIE team private communication). Therefore, we conclude that there is no significant offset between the new and the previous observations of WASP-103b. At $3\, \sigma$ we can also exclude RV variations higher than $151- 24  = 127$ m/s and lower than  $-119 - 24 = -143 $ m/s over 8 years. As outlined in Section~\ref{rvcompa}, this limit on the amplitude of the RV variation does not allow us to discard the bound scenario.

\subsection{Stellar parameters}

Thanks to the new data release of \textit{Gaia}, the stellar parameters of WASP-103A can be better constrained, which in turn allowed us to better constrain the mass and radius of the exoplanet WASP-103b.
 Using a modified version of the infrared flux method (IRFM; \citealt{Blackwell1977}), we determined the radius of WASP-103A via relationships between various stellar parameters recently detailed in \citet{Schanche2020}. We constructed the SED from stellar atmospheric models using the stellar parameters from SWEET-Cat \citep{Sousa2018} as priors, and subsequently attenuated the SED to account for reddening. The SED was further corrected for the companion using the calculated contamination estimate from the stellar parameters for the companion in \citet{Cartier2017} and reproduced in Table~\ref{parcompanion}. The corrected SED was then convolved with broadband response functions for the chosen bandpasses to obtain synthetic photometry which allowed us to compute the bolometric flux, and hence the radius, of the target. We retrieved broadband fluxes and uncertainties from the most recent data releases for the following bandpasses: {\it Gaia} G, G$_{\rm BP}$, and G$_{\rm RP}$; 2MASS J, H, and K; and {\it WISE} W1 and W2 \citep{Skrutskie2006,Wright2010,Gaia2021}.\ We also used the \textsc{atlas} catalogues \citep{Castelli2003} of model stellar SEDs. Within the IRFM, the distance used to convert the angular diameter of WASP-103A to the stellar radius was calculated from the {\it Gaia} EDR3 parallax with the parallax offset of \cite{Lindegren2021} being applied. Using a Markov chain Monte Carlo (MCMC) fitting approach, we estimated the stellar radius of WASP-103A to be $R_{\star}=1.716 \pm 0.119\,R_{\odot}$. This is larger than the previous estimates, that is $1.436 \pm 0.052$  \citep{Gillon2014} due to the greater distance to WASP-103 derived from the EDR3 parallax.

We derived both the stellar mass $M_{\star}$ and age $t_{\star}$ by employing two different sets of stellar evolutionary models, namely PARSEC v1.2S\footnote{\textsl{PA}dova \& T\textsl{R}ieste \textsl{S}tellar \textsl{E}volutionary \textsl{C}ode: \url{http://stev.oapd.inaf.it/cgi-bin/cmd}} \citep{Marigo2017} and CLES \citep[Code Liègeois d'Évolution Stellaire][]{Scuflaire2008}.
The input parameters we used were the stellar [Fe/H], $T_{\mathrm{eff}}$, and $R_{\star}$ to locate the star on the Hertzsprung-Russel Diagram (HRD) plus the projected rotational velocity $v\sin{i}$, which was plugged into the gyrochronological relation by \citet{Barnes2010c} to remove isochronal degeneracies; for further details, see \citet[\S 2.2.3]{Bonfanti2020}.
We performed a direct interpolation of the input parameters within pre-computed grids of PARSEC models thanks to the isochrone placement algorithm presented in \citet{Bonfanti2015} and \citet{Bonfanti2016}, obtaining the first pair of age and mass values. The second pair was inferred by directly computing the evolutionary track through the CLES code and then choosing the best-fit solution following the Levenberg-Marquadt minimisation scheme \citep{Salmon2021}. After checking the consistency of the two mass and age values following the $\chi^2$ criterion discussed in detail in \citet{Bonfanti2021}, we combined the respective distributions of age and mass together to obtain our final estimates of $M_{\star}$ and $t_{\star}$. The mass is in agreement with previous estimates, while the age is much better constrained \citep{Gillon2014,Delrez2018}. The final stellar parameters and the $1\,\sigma$ uncertainties are given in Table~\ref{stellarp}.

\begin{table}[h!]
\caption{Stellar parameters of WASP-103A. \label{stellarp}}
\begin{center}
\begin{tabular}{l  c }
\hline
\hline
Parameter & Value and uncertainty\\
\hline
Effective temperature \teff\ [K]  & 6013  $\pm$ 44 \\
Surface gravity \logg\ [g cm$^{-2}$]  & 4.24 $\pm$ 0.15\\
Iron abundance [Fe/H] [dex]  & 0.08 $\pm$ 0.04\\
Spectral type & F8V\\
Parallax*  $p$ [mas] & 1.8110 $\pm$ 0.1073\\
Distance to Earth $d$ [pc] & 552 $\pm$ 33 \\
Stellar mass $M_{\star}$ [M$_\odot$] &1.204 $\pm$ 0.046 \\
Stellar radius $R_{\star}$ [R$_\odot$] & 1.716  $\pm$ 0.119 \\
Stellar age $\tau$ [Gyr] &  5.2 $\pm$ 0.8\\
Stellar luminosity $L_{\star} $[L$_\odot$]  & 3.47$\pm$  0.49 \\ 
\hline
\hline
\end{tabular}
\tablefoot{\\
*Parallax from \textit{Gaia} EDR3  \citep{Gaia2021} using the formulation of \citet{Lindegren2021}. } \\
\end{center}
\end{table}

\subsection{Re-analysis of previous transits}
Given the very high signal-to-noise required to detect the tidal deformation, we have re-analysed high signal-to-noise transits of WASP-103 previously obtained with the Spitzer and Hubble space telescopes. These are combined with the 12 new transits obtained with CHEOPS in our final analysis.

\subsubsection{Spitzer observations}
\label{sec:spitzer}

We re-analysed the Spitzer archival data of WASP-103b which has already been published \citep{Kreidberg2018}.  We downloaded WASP-103b archival IRAC data from the Spitzer Heritage Archive (\url{http://sha.ipac.caltech.edu}).  The data consist of one full phase curve of WASP-103b at 4.5 $\mu$m (channel 2) and one at 3.6 $\mu$m (channel 1), both were obtained under program ID 11099 (PI L. Kreidberg) taken on 19 and 28 May 2015. The reduction and analysis of these datasets are similar to \citet{Demory2016}. We modelled the IRAC intra-pixel sensitivity \citep{Ingalls2016} using a modified implementation of the  BiLinearly-Interpolated Sub-pixel Sensitivity (BLISS) mapping algorithm (\citealt{Stevenson2012}). We used a modified version of the BLISS mapping (BM) approach to mitigate the correlated noise associated with intra-pixel sensitivity. In our photometric baseline model, we complement the BM correction with a linear function of the point response function (PRF) full width at half-maximum (FWHM). 

In addition to the BLISS mapping, our baseline model includes the PRF's FWHM along the $x$ and $y$ axes, which significantly reduces the level of correlated noise as shown in previous studies (e.g. \citealt{Lanotte2014}; \citealt{Demory2016}; \citealt{Demory2016b}; \citealt{Gillon2017}, \citealt{Mendonca2018}). Our baseline model does not include time-dependent parameters. Our implementation of this baseline model is included in an  MCMC framework already presented in the literature (\citealt{Gillon2012}). We ran two chains of 200,000 steps each to determine the phase-curve properties and obtained the best detrended transit light curves which are analysed together with HST and CHEOPS transits in Section~\ref{deformation}. From our BM+FWHM baseline model, we obtained a median root mean square (RMS) of 3450 ppm per 10.4s integration time at 3.6 $\mu$m and 4480 ppm with the same integration time at 4.5$\mu$m.

\subsubsection{HST observations}

We re-reduced the Hubble transit observations taken on 26-27 February 2015 and 2 August 2015 with HST Program 14050 which were originally published by \citet{Kreidberg2018}. The target was acquired in both forward and backward scanning direction using an exposure time of 103 s. We used the frames in the IMA format, each one containing 16 non-destructive reads \citep[NDR;][]{Deming2013}, which were pre-processed by the CALWFC3 pipeline\footnote{https://www.stsci.edu/hst/instrumentation/wfc3/software-tools/pipeline}, version 3.5.2.

Wavelength calibration, NDR operations, background subtraction, cosmic ray and bad pixels rejection, and correction for drifts were carried out following standard procedures, as described in \citet[][and references therein]{Bruno2018}. Then, we integrated the stellar spectra in the 1.115-1.625 $\mu$m wavelength range to obtain the band-integrated transits.
Following standard practice \citep[e.g.][]{Deming2013}, we rejected the first HST orbit of the transit obtained on 2 August 2015, which was at the beginning of the phase curve observation, and the first data point of every orbit for both transits.

We then used a method similar to \citet{Kreidberg2018} to remove the instrument systematics with the model described in \cite{Stevenson2014}, that is with a second-order polynomial and an exponential ramp,
\begin{equation}
    S(t) = C(1 + r_0\theta + r_1\theta^2)(1 - e^{r_2\phi + r_3} + r_4\phi),
\label{systmodel}
\end{equation}
where we fitted for $C$ and $r_{0-4}$, and $\theta$ and $\phi$ represent the planetary and \textit{HST} phase, respectively. It was also necessary to add a shift to the \textit{HST} orbital phase, $\phi = 2\pi [(t  -  \psi) \mod P_{HST}]/P_{HST}$, where $\psi= -0.045$~d is for the February 2015 visit and $\psi = -0.1$~d is for the August 2015 visit, respectively, and $P_{HST}$ is Hubble's orbital period.

 The systematics were fitted simultaneously with the transit model of a non-deformed planet using the model of \citet{Mandel2002} (implemented in \cite{Batman2015} software) and scipy's optimize.minimize function \citep[][and references therein]{scipy}. The best detrended transit light curves are analysed together with Spitzer and CHEOPS transits in Section~\ref{deformation}.The mid-times of each exposure were converted to BJD-TDB using the online applet based on the method of \citet{Eastman2010}.

\section{Period evolution of WASP-103b}
\label{ephemerides}

Using the procedure outlined in Section~\ref{dataanali}, we obtained the mid-transit times of the CHEOPS observations. These are given in Table~\ref{times}. In this table we also included the mid-transit times derived for the Spitzer and HST transits. We reiterate that the derived CHEOPS transit times obtained with our method are within $1\, \sigma$ of the ones derived using \pycheops\ showing that our detrending methods are robust.

We combined our derived mid-transit times (12+4) with the 32 previously published mid-transit times of WASP-103 which were presented in Table~3 of \citet{Maciejewski2018}, some of which are reanalyses of previously published values \citep{Gillon2014,Southworth2015,Delrez2018,Turner2017, Lendl2017}. We also added the four transit times subsequently presented in \citet{Patra2020}. Therefore, in total we have 52 mid-transit times of WASP-103b spanning seven years.

For parameter inference, we used \emcee\ as in Section~\ref{dataanali}. We compared a linear ephemerides (constant period) model with a quadratic ephemerides (constant derivative period) model. We included a multiplicative jitter parameter in our analysis.

For the constant period model, we obtained $T_0\,=\,2457511.944458^{+0.000049}_{-0.000048}\, (BJD_{TDB} ) $ and $P = 0.925545485 \pm 0.000000049\,$days. The BIC of this fit is  79.8. We found a jitter of 1.18.

We considered a constant derivative period model with the following form \citep[e.g.][]{Maciejewski2020}:
\begin{equation} 
\label{quadratic}
T_{mid} =  T_0 +  P \times E + P \dot{P}\times \frac{ E(E-1)}{2},
\end{equation} 
where $E$ is the transit epoch and  $\dot{P}$ is the period derivative.

For this model we derived  $T_0\,=\,2457511.944344 \pm 0.000075 (BJD_{TDB} ) $,   $P = 0.9255453 \pm 0.000000089$ days, $  \dot{P}  =  3.5 \pm 1.8 \times 10^{-10} $ days/day, and jitter $= 1.15$. The jitter is slightly lower than for the linear model. We found a hint of an increasing orbital period, contrary to what was expected if tidal decay was dominating the orbital evolution of the system. The period derivative was found to be positive at $2.1\, \sigma$ which is not significant. The BIC of the quadratic model is 78.05, giving a difference of BIC in favour of the quadratic model of only 1.8. Therefore, according to the BIC, the added complexity of the quadratic model is not strongly justified and the linear ephemerides is preferred.

\begin{figure} 
\centering 
\includegraphics[width=0.9\columnwidth]{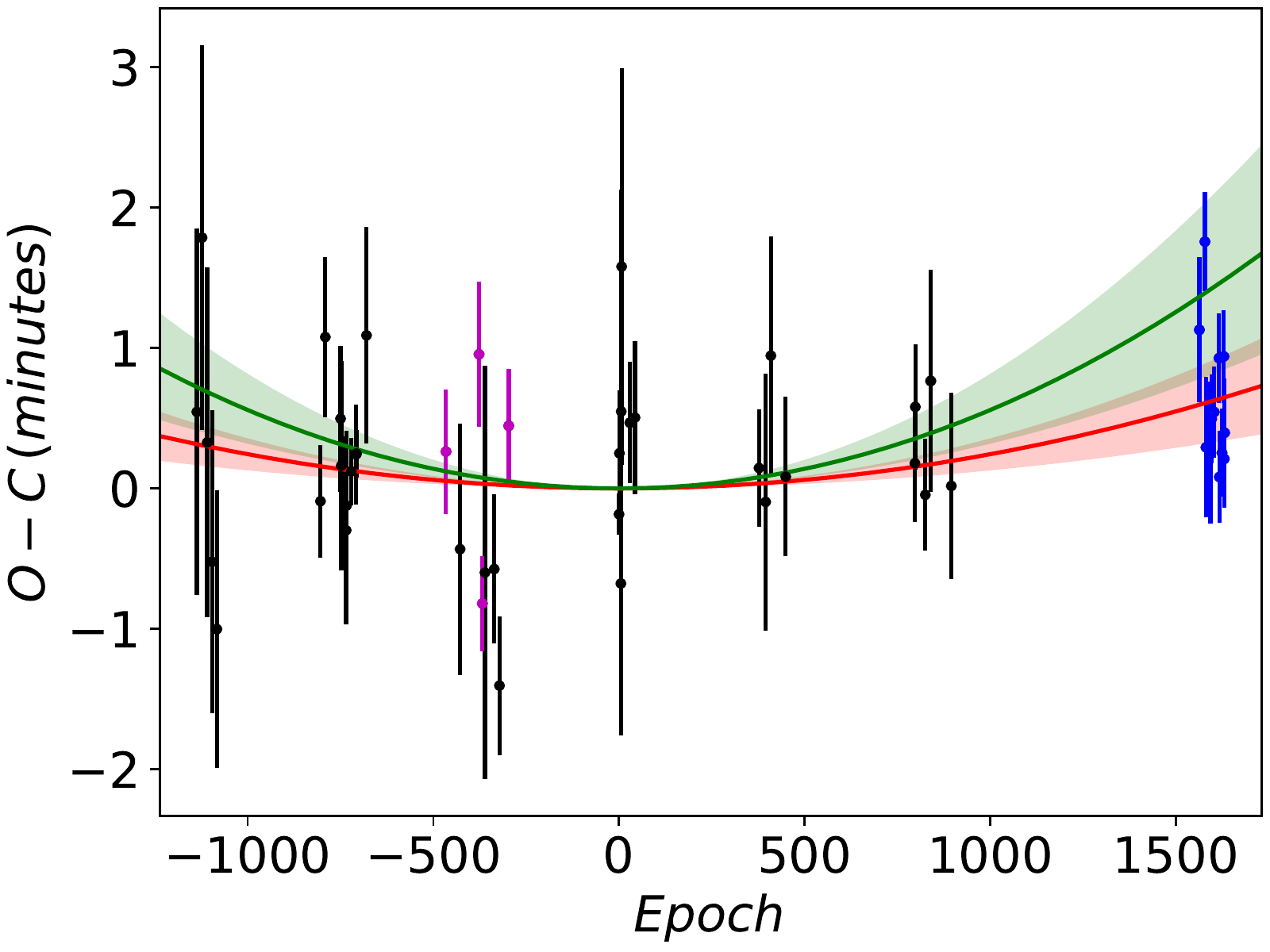}
\caption{Derived mid-transit times of WASP-103b after removing a linear ephemerides. The CHEOPS data are shown in blue, the re-reduced HST data and Spitzer are shown in magenta, and the previously published times are shown in black. The previously published quadratic ephemerides \citep{Patra2020} is shown in green with the $1\, \sigma$ uncertainty limit, while our new result is shown in red. \label{o_c}} 
\end{figure}

Under assumption that the period variation observed is due to tidal decay (i.e. the period is actually decreasing and the variation seen is due to statistical uncertainties), we can derive a lower limit to the tidal dissipation parameter using the following equation \citep[e.g.][]{Maciejewski2020}: 
\begin{equation}
 Q'_* = \frac{27\pi}{2}\frac{M_p}{M_*} \left(  \frac{R_*}{a} \right)^5\frac{1}{ \dot{P} },
\end{equation} 
where $M_p$ and $M_*$ are the planetary and stellar masses, respectively, $R_*$ is the stellar radius, and $a$ is the semi-major axis of the planet's orbit. We derived a lower limit on the period derivative to be $-1.3 \times 10^{-10} $ days/day at the 99.7\% confidence interval. This implies that the tidal dissipation parameter is higher than $1.6 \times 10^6$ at $3\, \sigma$, corresponding to a 99.7\% confidence interval if we assume that only tidal decay affects the period derivative. This limit is more than an order of magnitude higher than previous studies that found  $ Q'_*  > 1.1 \times 10^5$ at 95\% \citep{Patra2020} for WASP-103b. At 95\% confidence, our results allow us to exclude a negative period derivative. The quadratic fit to the derived transit times is given in Figure~\ref{o_c}. We found a period derivative that is smaller than the previous estimation \citep{Patra2020}, although the higher precision results in a higher significance for being positive.

Figure~\ref{o_c} shows that the first two CHEOPS transits have a slightly late mid-transit time compared with the other observed CHEOPS transit times, although consistent within the errors. This is probably due to the difficulty in detrending CHEOPS data when the duration of the visit is shorter than three CHEOPS orbits. It is also known that transits with poorly covered ingress or egress can lead to biases in the derived transit times \citep{Barros2013}. To test if this could influence our results we repeated the linear and quadratic model fits excluding the first two transits. We found no significant differences in the derived model parameters or model comparison. We also repeated the fits using the transit times derived with \pycheops\ instead of the ones derived with the multi-dimensional GP and found the same results.

Although it is likely that the observed period variation is due to statistical dispersion and that the orbit is decaying due to tides, it is interesting to explore other factors that could affect the orbital evolution of this system. In the next subsections, we explore scenarios that could explain an increase in the orbital period in case it becomes significant after future observations.

 \begin{table} 
\caption{Derived mid-transit times of WASP-103b.  \label{times} }
\begin{tabular}{lccc} 
\hline 
\hline 
Epoch & time &  $\sigma_{+} $ &  $\sigma_{-} $ \\ 
\hline
  & BJD TDB -2450000  (days) &(days) &(days) \\
 -466  &7080.64040 &    0.00033 &0.00028\\
 -377  &7163.01442 &    0.00036 &0.00036\\
 -368  &7171.34309 &    0.00024 &0.00024\\
 -297  &7237.05769 &    0.00027& 0.00030\\
1563 &  8958.57247 &    0.00037 &       0.00035 \\
1578 &  8972.45609 &    0.00024 &       0.00025\\
1581 &  8975.23171 &    0.00036 &       0.00034 \\
1593     &      8986.33822 &    0.00039 &       0.00031 \\
1596     &      8989.11503 &    0.00024 &       0.00020 \\
1603     &      8995.59388 &    0.00022 &       0.00023 \\
1616     &      9007.62623 &    0.00023 &       0.00022 \\
1617     &      9008.55119 &    0.00022 &       0.00023 \\
1624     &      9015.03013 &    0.00022 &       0.00022 \\
1629     &      9019.65833 &    0.00023 &       0.00023 \\
1630     &      9020.58337 &    0.00025 &       0.00023 \\
1631     &      9021.50904 &    0.00025 &       0.00029 \\
\hline 
\hline 
\end{tabular}
\tablefoot{\\
The first four entries refer to the HST and Spitzer transits, while the last 12 were derived from the new CHEOPS observations.}
\end{table}

\subsection{RV acceleration due to a companion}
\label{rvcompa}
The existence of a companion of WASP-103 could lead to RV acceleration which would produce transit timing variations due to a change in the light travel time.
Assuming a quadratic ephemerides and that the observed period derivative is due to the Doppler effect of a line-of-sight acceleration ($\dot{P}_{RV}$), we can derive the line-of-sight acceleration ($a_r$) using the following: 
\begin{equation} 
\label{doppler}
\dot{P}_{RV} = \frac{a_r\, P}{c}
,\end{equation} 
where $c$ is the speed of light. We obtained $a_r = 0.113 \pm 0.058$ m/s/day.

During the eight years since the discovery of WASP-103b until now, this would imply an RV variation of $330\pm168$ m/s. This is higher than the RV offset that we measured in Section~\ref{RV}, but it is still compatible within the errors.
 
We can also calculate the expected acceleration from the visual companion of WASP-103 if it is bound (discussed in Section~\ref{companion}) using Newton's law:
\begin{equation} 
\label{newton3}
a = \frac{GM_{comp}}{d^2}  
,\end{equation} 
where $M_{comp}$ is the mass of the visual companion star and $d$ is the separation between the two stars. In this case, where the companion is observed to be spatially separated from WASP-103, we need to account for the projection of the acceleration in the line-of-sight given that only the radial component of the acceleration results in a variation of the observed period. If the angle between the line-of-sight and the companion is $\theta$, then $a_{rad} = a \cos \theta$ and $d = \delta/ \sin\theta$. Where $\delta$ is the projected separation of the star and the companion that we previously derived (Section~\ref{companion}) as $134 \pm 8 $ au. Hence,
\begin{equation} 
\label{newton3rad}
a_{rad} = \frac{GM_{comp}}{\delta^2} (\sin\theta)^2 \cos \theta
.\end{equation} 

The maximum value of equation \ref{newton3rad} is obtained for $\cos \theta = \frac{1}{\sqrt{3}}$. Assuming  $M_{comp} = 0.721\pm  0.024 $ \Msun (Table~\ref{parcompanion}), we can set an upper limit on  $a_{rad}  \leqslant   0.00796 \pm 0.00095 $  m/s/day. This corresponds to an RV acceleration of $23.6 \pm 2.7$ m/s in 8 years. This is compatible with the RV offset that we measured between the new CORALIE observations and the previously published observations (Section~\ref{RV}). The difference between the acceleration expected if the transit timing variations are due to acceleration from a bound companion and the acceleration from the observed visual companion is $0.105 \pm 0.057$ m/s/day. Hence, the acceleration produced by the visual companion is probably not enough to produce the period derivative estimated with the quadratic model. However, we cannot exclude it at more than $2\, \sigma$.

Long-term RV monitoring of WASP-103b and the next \textit{Gaia} DR3 will allow us to better constrain the existence of possible bound companions to WASP-103 and correct the line-of-sight acceleration light travel time, allowing us to better constrain the tidal dissipation parameter. The hypothesis that the visual companion observed by lucky imaging and AO observation is responsible for the transit timing variations and the offset in CORALIE RVs cannot be completely rejected. However, the absence of this star in our new lucky imaging observation and the fact that the predicted acceleration by this star is $2\, \sigma$ lower than required to match the observations, suggests that other mechanisms would be required to explain an increase in the orbital period of the planet.

\subsection{Applegate effect}

Eclipse times of binary stars have been shown to vary due to variations in the quadrupole moment of the
stars driven by stellar activity. Low mass stars with convective outer layers have variations of their quadrupole moment due to a distribution of angular momentum driven by stellar activity cycles. The change of the quadrupole moment of the star leads to quasi-periodic variations of the eclipse times of the companion over timescales of years to decades. This effect has been measured in many eclipsing binaries and is known as the Applegate effect \citep{Applegate1992}. 
Another explanation for the observed period changes in binaries was proposed by \citet{Lanza1998}. In this case, the Applegate effect would be due to a cyclic transformation of rotational kinetic energy into magnetic energy and back to rotational kinetic energy. If the Applegate effect is detected, it would allow us to probe the nature of the dynamo mechanism of low mass stars \citep{Lanza1998}.

Since exoplanets host low mass stars with some dynamo activity, it is expected that the Applegate effect is also present in exoplanet systems; although, it has never been observed. \citet{Watson2010} estimated the transit time variations due to the Applegate effect for a few transiting exoplanets. They show that for stellar dynamos with timescales of 11 years, the Applegate effect is less than $5$ seconds for most exoplanet host stars. However, for stellar dynamos with longer timescales, the effect can reach a few minutes. Using their equation 13 and assuming the stellar parameters given in Table~\ref{stellarp}, the semi-major axis of the orbit $a = 0.01985$ au, the observation time span $T=7$ years, and estimating the stellar angular rotation velocity of WASP-103 from the $v\sin{i}$ given in \citet{Gillon2014} ($ 10.6\,$ \kms\.), we conclude that the Applegate effect in WASP-103b would produce transit timing variations $\leq 38$ seconds over the time span of the available observations. Assuming the quadratic ephemerides, we found that at the mid-epoch of CHEOPS observations,  the measured transit time, is  $1.69 \pm 0.81$ minutes later than what would be expected by a linear ephemerides. Therefore, this is higher than the expected timing variations from the Applegate effect, although in agreement at $1.3\,\sigma$. Hence we conclude that the Applegate effect could be affecting the measured transit times of WASP-103b.

\subsection{Apsidal precession}
If a planet's orbit is slightly eccentric, then its orbit would be apsidally precessing. For hot Jupiters, the precession timescale is expected to be decades. In this case, there is a long-term oscillation of the apparent period. Modelling the period variation would allow us to determine the planet Love number and constrain its internal structure. 
For WASP-103b, a zero eccentricity was assumed by \citet{Gillon2014} and favoured by the analysis of \citet{Delrez2018} and our own analysis using the new RV measurements presented in Section~\ref{RV}. 
 
Nevertheless we attempted to fit the times of WASP-103 with a transit timing model assuming apsidal precession \citep{Patra2017}:
\begin{equation} 
\label{apsidal}
T_{mid} =  T_0 +  P_s \times E - \frac{e\, P_a}{\pi}   \times \cos\omega,
 \end{equation} 
where,
\begin{equation} 
P_s=Pa \left(  1 - \frac{ d\omega/dE}{2\pi} \right)
 .\end{equation} 
 We found that the apsidal precession model is a poor fit to the transit times with a BIC =  82.7. This is due to the two extra parameters compared with the quadratic model that is already not justified by the BIC compared with the linear model. However, we obtained physical values for the fitted parameters. We obtained $\frac{ d\omega}{dE} = 1.10^{+1.9}_{-0.63} \times 10^{-3}$ rad, $e =0.00054^{+0.0055}_{-0.000006} $, and $\omega = 0.49^{+0.57}_{-0.96}$ rad. 
 
 The most important contributions to the apsidal precession rate of hot Jupiters are those coming from the tides raised by the star on the planet and from the rotation of the planet \citep{Ragozzine2009}.
 Assuming synchronous rotation, the leading term in the expression of this precession rate at low eccentricity is as follows (see, e.g. eqs (6)+(10) of \citealt{Ragozzine2009}):
 \begin{equation} 
 \label{prate}
\left(  \frac{ d\omega}{dE} \right)_{T+R} =16 \pi (h_f- 1) \frac{M_*}{M_p}  \left(  \frac{ R_p}{a} \right)^5 
 .\end{equation} 
Using our fitted value for the precession rate of $ 1.10^{+1.9}_{-0.63} \times 10^{-3}$ rad, we can estimate the Love number to be $h_f =  1.35 \pm 0.43$ which is compatible with our estimate from the deformation of the light curve (Section~\ref{deformation}).

 Current observational constraints on the eccentricity cannot rule out such a small value $\sim 0.00054$. However, due to the short circularisation timescale, the eccentricity of WASP-103b is expected to be zero unless there is an external perturbation. For example, a planetary companion can excite the eccentricity of WASP-103b. The eccentricity can also be excited by gravitational perturbations from the star's convective eddies as proposed by \cite{Phinney1992}.

 Therefore, we cannot completely rule out that apsidal precession is affecting the transit times of WASP-103b given our current constraints on the eccentricity of the planet although this is, a priori, not expected. Future monitoring of the transit times of WASP-103b can disentangle apsidal precession from tidal decay since for apsidal precession the variations are sinusoidal. The times of occultation can also be used to disentangle both scenarios because in the apsidal precesion, these are anti-correlated with the times of transit.

\section{Tidal deformation analysis}
\label{deformation}

As mentioned above, WASP-103b is the exoplanet with the highest expected deformation signature due to its large radius and close proximity to its host star. We attempted to measure the deformation and tidal Love number of WASP-103b, combining the 12 new high-precision transits obtained with CHEOPS with two HST transits and two Spitzer transits (3.4 and 4.5 microns). To model the tidal deformation, we used the implementation of \citet{Akinsanmi2019} based on the parametrisation of \citet{Correia2014}. This implementation uses the \project{ellc} transit tool \citep{Maxted2016} and it is also freely available. The model parameters are the normalised separation of the planet ($a/R_{\star}$), the impact parameter ($b$), the Love number ($h_f$), the logarithm of the planet-to-star mass ratio multiplied by the sine of the inclination ($\ln Q_m = \ln \left( \frac{M_p}{M_*} \sin{inc} \right) $), and, for each filter, the planet-to-star radius ratio ($R_{V}/R_{\star}$) and the power-2 limb darkening (LD) coefficients ($c$  and $\alpha$). Following equation~\ref{hf}, in this ellipsoidal model, the radius of the planet is parameterised by the volumetric radius $R_{V} =\sqrt[3]{r_1r_2r_3}$. The LD coefficients were parameterised according to \citet{Kipping2013b} and \citet{Short2019} to minimise the correlations between them and to avoid non-physical solutions.

The priors for each parameter are given in Table~\ref{prior_ellipsoid}.  For the shape parameters, we used uniform uninformative priors instead of normal distributions based on previous data because the previous results were obtained assuming sphericity, which impacts the derived shape parameters. We assumed the period and mid-transit times derived in Section~\ref{ephemerides}. We included a multiplicative jitter term for CHEOPS, HST, and Spitzer channel 1 and channel 2 to account for any underestimation of the uncertainties. For each light curve, we corrected the contamination due to the visual companion star (see Section~\ref{companion}), assuming the stellar parameters given in Table~\ref{parcompanion} and the respective filter transmission functions.

For the parameter inference, we used the nested sampling algorithm implemented in  \project{Dynesty} \citep{Speagle2020,Higson2019,Skilling2012,Skilling2004} which provides posterior estimates and also the Bayesian evidence useful for model comparison. We fitted the tidal deformation parameterised by $h_f$ and compared it with a spherical model ($h_f =0$). The comparison of the models illustrates biases in the derived shape parameters if the deformation is ignored. The derived transit parameters are given in Table~\ref{results} for the spherical and the ellipsoidal model. The derived jitter parameters for the  ellipsoidal model are 1.00, 1.11, 1.21, and 1.09 for Spitzer channel 2, Spitzer channel 1, CHEOPS, and HST,  respectively, showing that our errors are robust and the detrending was successful. The derived jitter parameters for the spherical model are similar to the  ellipsoidal model.

\begin{table} 
\caption{Priors for the fitted transit parameters. \label{prior_ellipsoid} }
\begin{tabular}{lc} 
\hline 
\hline Parameter & Prior  \\ 
$a/R_{\star}$  &  $\mathcal{U}( 2.5,3.5)$   \\ 
$b$  &  $\mathcal{U}( 0,1)$   \\ 
$h_f$ &  $\mathcal{U}( 0,2.5)$   \\ 
$\log Q_m$ &  $\mathcal{N}( -6.7581, 0.0534) $ *  \\ 
$R_p/R_{\star}$  each instrument& $\mathcal{U}( 0.05,1.5)$ \\ 
&\\
$c$ CHEOPS&  $\mathcal{N}(  0.7045, 0.0147)$  \\ 
$\alpha$ CHEOPS &  $\mathcal{N}(0.7670, 0.0199)$  \\ 
$c$ HST&  $\mathcal{N}( 0.5714, 0.0218)$   \\ 
$\alpha$ HST &  $\mathcal{N}( 0.4285, 0.0175)$  \\ 
$c$ Spitzer 1&  $\mathcal{N}(0.2772, 0.0085)$   \\ 
$\alpha$ Spitzer 1 &  $\mathcal{N}( 0.4730, 0.0229)$  \\ 
$c$ Spitzer 2&  $\mathcal{N}( 0.2280, 0.0067)$   \\ 
$\alpha$ Spitzer 2 &  $\mathcal{N}( 0.483, 0.021)$  \\ 
\hline 
\hline 
\end{tabular} 
\tablefoot{\\ * Derived from the RV analysis in Section~\ref{RV}.} \\
\end{table}

\begin{table*}[h!]
\caption{Derived transit parameters of WASP-103b for an ellipsoidal planet model and a spherical model.  \label{results} }
\begin{center}
\begin{tabular}{l  c c}
\hline
\hline
Parameter             & Spherical Model fit  (S)                         & Ellipsoidal Model fit (E) \\
\hline
Love number $h_f$                & --                                                              & $1.59^{+0.45}_{-0.53}$ \\
ln Qm                    & --                                                              & $-6.761 \pm 0.049$ \\ 
 $a/R_{\star}$        & $ 2.9975^{+0.0061}_{-0.0011}$        & $3.0038^{+0.0046}_{-0.0070}$ \\
 $b$       & $0.066^{+0.049}_{-0.039}$                 & $0.044^{+0.040}_{-0.027}$ \\
 $M_{p}/\Mjup$   & $1.464 \pm 0.096 $     & $1.460 \pm 0.089$ \\
&\\
$R_{V}/R_{\star}$  Spitzer 2 & $0.12245 \pm 0.00086 $     & $0.1297 \pm 0.0028$ \\
$R_{V}/R_{\star}$ Spitzer 1 & $0.11928 \pm 0.00059$        & $0.1257 \pm 0.0024$ \\
$R_{V}/R_{\star}$      HST       & $0.11639 \pm  0.00018$     & $0.1222 \pm 0.0021$ \\
$R_{V}/R_{\star}$   CHEOPS  & $0.11588 \pm 0.00020$      & $0.1215 \pm 0.0020$ \\
$R_{V}$ [R$_{Jup}$] Spitzer 2 & 2.044 $\pm$ 0.142          &  2.165 $\pm$ 0.155\\
$R_{V}$ [R$_{Jup}$] Spitzer 1 & 1.992 $\pm$ 0.138          & 2.097 $\pm$ 0.150\\
$R_{V}$ [R$_{Jup}$] HST            & 1.943 $\pm$ 0.135           & 2.039 $\pm$ 0.144 \\
$R_{V}$ [R$_{Jup}$] CHEOPS    & 1.935 $\pm$ 0.134           & 2.027 $\pm$ 0.145\\
$\rho_{p}$ [$\rho_{Jup}$] Spitzer 1  & $0.171^{+0.043}_{-0.033} $  & $0.144^{+0.037}_{-0.028} $   \\
$\rho_{p}$ [$\rho_{Jup}$] Spitzer 1  & $0.185^{+0.046}_{-0.035} $  & $0.158^{+0.041}_{-0.031} $   \\
$\rho_{p}$ [$\rho_{Jup}$]  HST  & $0.199^{+0.049}_{-0.038} $            & $0.173^{+0.043}_{-0.033} $   \\
$\rho_{p}$ [$\rho_{Jup}$]  CHEOPS  & $0.202^{+0.050}_{-0.038} $    & $0.175^{+0.044}_{-0.034} $   \\
&\\
\hline 
\hline 
\end{tabular} 
\\
\end{center}
\end{table*}

 \begin{figure*}[ht!]
\centering 
\includegraphics[width=2.0\columnwidth]{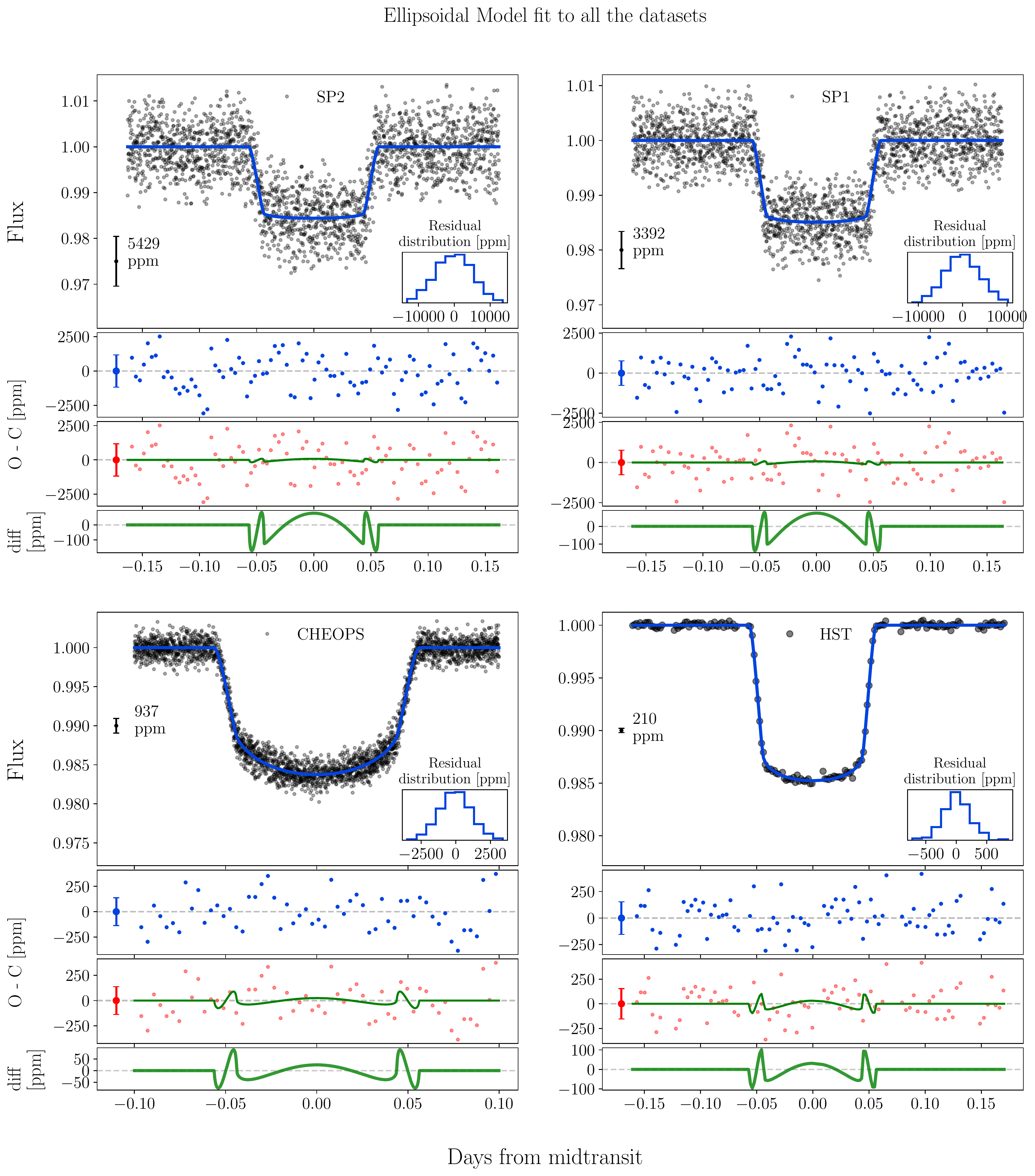}
\caption{Time folded transit light curves of WASP-103b obtained with Spitzer 1, Spitzer 2, CHEOPS, and HST. The CHEOPS transit light curve is a combination of 12 individual transits, while the HST light curve is a combination of two transits. The best fit ellipsoidal transit model is shown in blue. We also plotted the residuals of the best fit ellipsoidal model (blue) and the best fit spherical model (red) binned to 5 minutes. On the latter, we overplotted the signature of the deformation (green) which is the difference between the best fit spherical model and the best fit ellipsoidal models. For clarity, we replotted a zoom of the signature of the deformation in the bottom panel for each filter. We also show the mean uncertainties of the original data points and of the binned residuals. \label{deforfig}} 
\end{figure*}

 We overplotted the best model that accounts for tidal deformation on the time-folded light curves of WASP-103b taken with Spitzer 2, Spitzer 1, CHEOPS, and HST in Figure~\ref{deforfig}. We also show the residuals of the spherical model and overplotted the difference between the best fit spherical model and the best fit ellipsoidal model. As shown by \citet{Akinsanmi2019}, this is the measurable signature of the deformation of a planet in a transit light curve. This signature has two components. The first one is the signature of the oblateness ($r_2 > r_3$ ) resulting in an oscillation in the residuals of the flux during ingress and egress  \citep{Seager2002, Barnes2003}. The second one rises from $r_1 > r_2$ due to the change of the projected area of the ellipsoidal planet as it rotates synchronously with its orbit. This results in a bump that has its maximum at the minimum of the projection which is the middle of the transit \citep{Correia2014}. A schematic view of the geometry of how the deformation changes a transit light curve is given in Figure A.1 of \citet{Correia2014}. The change in the amplitude of signature of the deformation with the wavelength of the observations due to the change in the limb darkening and the larger planetary radius at longer wavelengths, as it can be seen in Figure~\ref{deforfig}, is noteworthy. This prevented us from phase folding all light curves and signatures so that our results could be visualised better.

\begin{figure*}[ht!]
\centering 
\includegraphics[width=2.0\columnwidth]{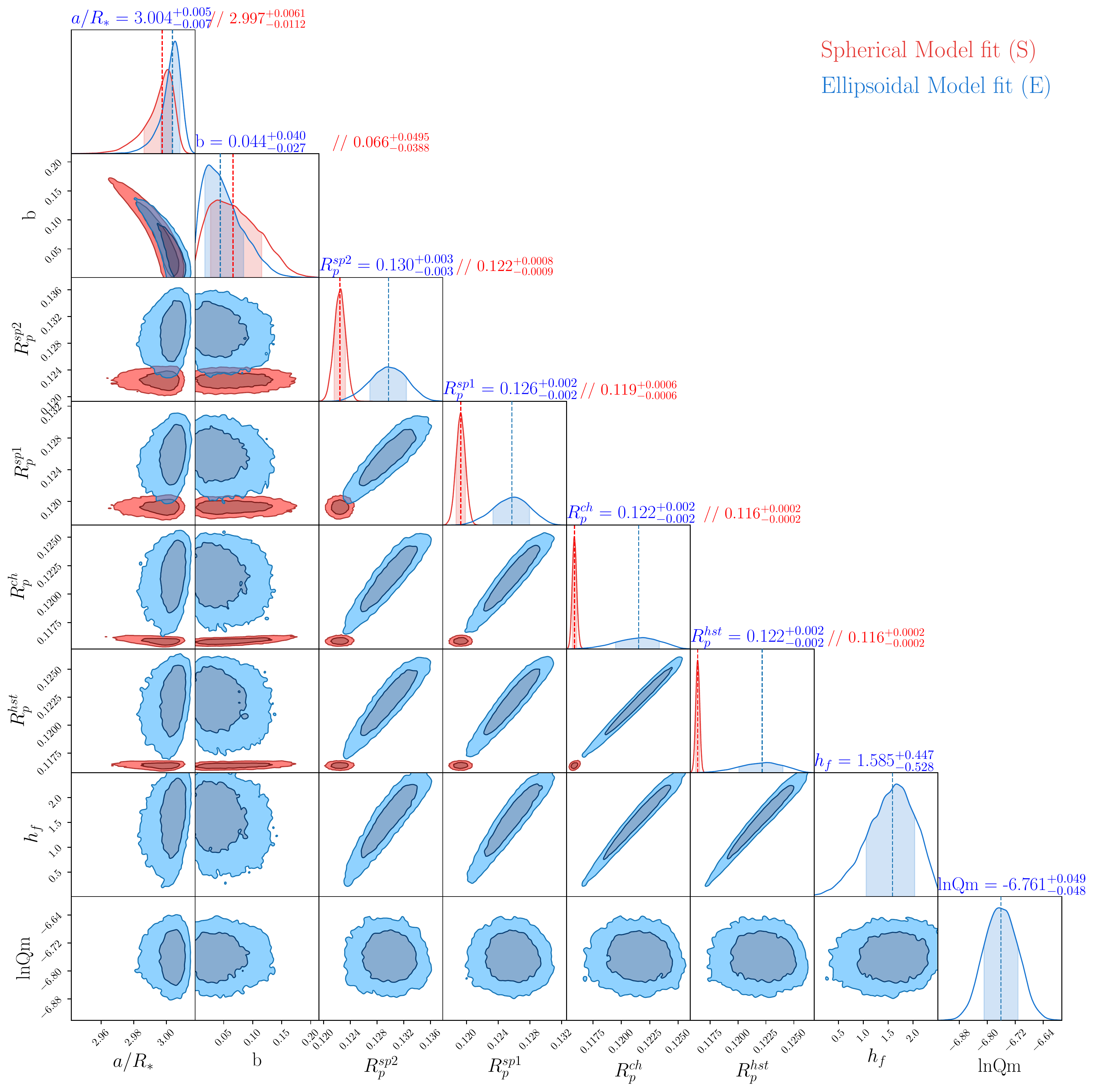}
\caption{Derived correlation plots and posterior probability distributions of the transit parameters of WASP-103b for the spherical (red) and ellipsoidal model (blue). The vertical lines show the median of the distributions and the shaded area shows the 68\% confidence intervals. We show the $1\,\sigma$ (dark blue and dark red) and $2\,\sigma$ (light blue and light red) contours.  We obtained a $3\, \sigma$ detection of the Love number. The parameter distributions also clearly show that the ellipsoidal model is not as well constrained as the spherical model due to strong correlations between the Love number and the radius ratio. For the ellipsoidal model, the radius ratio refers to the volumetric radius. The superscripts sp1, sp2, ch, and hst refer to the two Spitzer channels, CHEOPS, and HST, respectively. \label{chains}} 
\end{figure*}

 In Figure~\ref{chains} we show the correlation plots and the posterior probability distributions for the derived transit parameters of WASP-103b. As expected, there is a large correlation between the Love number and the radius ratio for the ellipsoidal model that leads to a larger uncertainty of the parameters of this model.  For simplicity, we do not show the distribution of the LD parameters and the jitter parameters because their shape is very well approximated by a Gaussian and they are very similar for the two models.

We derived the radial Love number of WASP-103b to be $h_f =1.59^{+0.45}_{-0.53}$. This is the first time that a $3\, \sigma$ detection of the Love number has been achieved for an exoplanet directly from the analysis of the deformation of the transit light curve. To obtain this result we combined the datasets from the three instruments: CHEOPS, HST, and Spitzer. To show the importance of each data set, we fitted the Spitzer and HST light curves separately and together. These results are given in Table~\ref{datasets} and show that the addition of the CHEOPS data was necessary to obtain a $3\, \sigma$ detection. It also justifies that the signature is not evident by eye in Figure~\ref{deforfig} for any individual datasets. We conclude that the Love number of WASP-103b is similar to the Love number measured for Jupiter ($1.565\pm 0.006$ -- \citealt{Durante2020}), suggesting a similar internal structure despite the much larger radius and much higher levels of irradiation for this exoplanet. The derived Love number of WASP-103b is higher than the one estimated for HAT-P-13b by \cite{Batygin2009}. This new measurement of the Love number can be used to lift the degeneracy of internal composition models \citep{Baumeister2020} and allows the derivation of the core mass of WASP-103b similarly to what was done for HAT-P-13b \citep{Kramm2012,Buhler2016}.

\begin{table} 
\caption{Comparison of the derived Love number for the individual instruments and their combination. \label{datasets}}
\begin{tabular}{lccc} 
\hline 
\hline Data set & Love number & Significance & Bayes Factor \\ 
SP2, SP1 & $1.36^{+0.71}_{-0.79}$ & 1.7 $\sigma$  & 1.7 \\
HST & $0.99^{+0.68}_{-0.59}$  & 1.7$\sigma$ & 0.71\\
CHEOPS & $1.74^{+0.69}_{-0.49}$ & 2.5 $\sigma$ & 6.1 \\
HST, SP2, SP1 & $1.16^{+0.64}_{-0.63}$ & 1.8 $\sigma$ &  1.0\\
All data  & $1.59^{+0.45}_{-0.53}$ & 3.0 $\sigma$ & 9.1 (17$^*$) \\
\hline 
\hline 
\end{tabular}
\tablefoot{\\
$^*$ Corrected value of the Bayes factor as explained in Section~\ref{signi}.}
\end{table}

 We found that the volumetric radius derived with the ellipsoidal model is 5-6\% bigger than the radius estimated with the spherical model. Therefore, not accounting for deformation biases the derived planetary radius and hence the planetary density ($\sim 14\%$) and composition. This is the first time that this bias that was predicted by \citet{Burton2014} and  \citet{Correia2014} has been directly measured.
 The large tidal deformation in ultra-hot Jupiters affects their phase curve observations and consequently their atmospheric characterisation. Previous phase curve measurements of WASP-103b \citep{Delrez2018, Lendl2017, Kreidberg2018}  have corrected tidal deformation using theoretical estimations   \citep[e.g.][]{Budaj2011,Leconte2011} that assume an interior structure for the planet. Our measurement of the Love number will allow an assumption-free correction based on direct observations. This will allow a more accurate estimation of the day-side and night-side temperatures from phase curve observations. It is also possible that neglecting to account for the deformation of WASP-103b could affect the interpretation of its transmission spectra \citep{Lendl2017}.
 
 \subsection{Assessing the significance of the detection}
 
\label{signi}

 One way to assess the significance of the detection is to perform model comparison - probability of one hypothesis versus another. Bayesian model comparison  requires computing the odds ratio between two hypotheses \citep[e.g.][]{Diaz2014}. The odds ratio is the multiplication between the prior odds and the Bayes factor (ratio of the Bayesian evidences). The prior odds are the {a priori} probability of each model. Given the strong tidal forces WASP-103b is subjected to by its host star, theoretically, we know that the planet has to be deformed. Hence, the prior probability of the spherical model is zero which implies that the odds ratio in favour of the ellipsoidal model is infinity and renders the Bayes factor irrelevant. Nevertheless, we estimated the Bayes factor of the ellipsoidal compared with the spherical model using the evidence computed with \project{Dynesty}. We found the Bayes factor (ratio of the Bayesian evidences) of the ellipsoidal compared with the spherical model to be 9.1, giving positive evidence for the ellipsoidal model \citep{Kass1995}. However, the Bayes factor penalises more complex models which is incorrect in our case since, as mentioned above, the planet is expected to be deformed and not accounting for deformation significantly biases the derived transit parameters, especially the planetary radius. To correct the penalisation of the extra parameters, we fitted an ellipsoidal model with a fixed value of $h_f$ and ln Qm, corresponding to the best fit model. We found the Bayes factor rises to 17.2. Hence, according to this corrected value for the Bayes factor, the ellipsoidal model is 17 times more probable than the spherical model meaning that the data show positive evidence for the deformation model. 

Furthermore, in our case we do not need to compare two hypothesis but we need to access the detectability of a measurement and hence we should use parameter inference instead of model comparison. Hence, instead of answering the question of whether the planet is deformed we answer the question of how much the planet is deformed. This latter question is best assessed by the analysis of the posterior probability distribution of $h_f$ which measures the deformation rather than by model comparison. Since we found that the distribution of $h_f $ does not include the spherical model ($h_f =0$) at $3\, \sigma$, we conclude that the deformation was detected. Using the posterior distribution of $h_f$, we can compute more accurate limits on the detection given that the distribution is not completely Gaussian. We found that $h_f$ is higher than $0.18$ at the $99.7$ confidence limit ($3\, \sigma$) and $h_f$ is higher than $0.03$ at the $99.95$ confidence limit ($3.5\, \sigma$). Hence, the  detection is slightly higher than $3\, \sigma$.

\subsection{Impact of limb darkening}

The model of the limb darkening can affect the measurement of the Love number \citep{Akinsanmi2019, Hellard2019}. Despite several studies on the best way to model the LD in exoplanet light curves \citep{Csizmadia2013,Howarth2011}, consensus still eludes us as it appears that the best model might depend on the quality of the data being analysed  \citep[e.g.][]{Espinoza2015}. Of the several LD parametrisations, the non-linear law \citep{Claret2000} is usually regarded as the best description of the stellar intensity profile \citep{Howarth2011}; however, when fitting the parameters in the transit light curves, the correlations between the four parameters can lead to non-physical models. Recently, the power-2 law \citep{Hestroffer1997} has been shown to be a good balance between a small number of parameters and being a good approximation of the stellar intensity profiles \citep{Morello2017}.  Therefore, it has gained much interest helped by a faster algorithm \citep{Maxted2019}. The most commonly used LD law is the quadratic law \citep{Kopal1950} due to its relative simplicity, fast implementation, and the existence of several parametrisations to minimise correlations between the two parameters  \citep[e.g.][]{Kipping2013b}. In addition to the choice of the parametrisation, it is also unclear if it is best to fix the LD coefficients to theoretical values based on stellar models or directly fit the LD in the light curves. The best approach depends not only on the precision of the light curves \citep[e.g.][]{Espinoza2015}, but also on the geometry of the system  \citep[e.g.][]{Howarth2011} and on whether the star is active \citep{Csizmadia2013}.

\begin{figure}[ht] 
\centering 
\includegraphics[width=0.9\columnwidth]{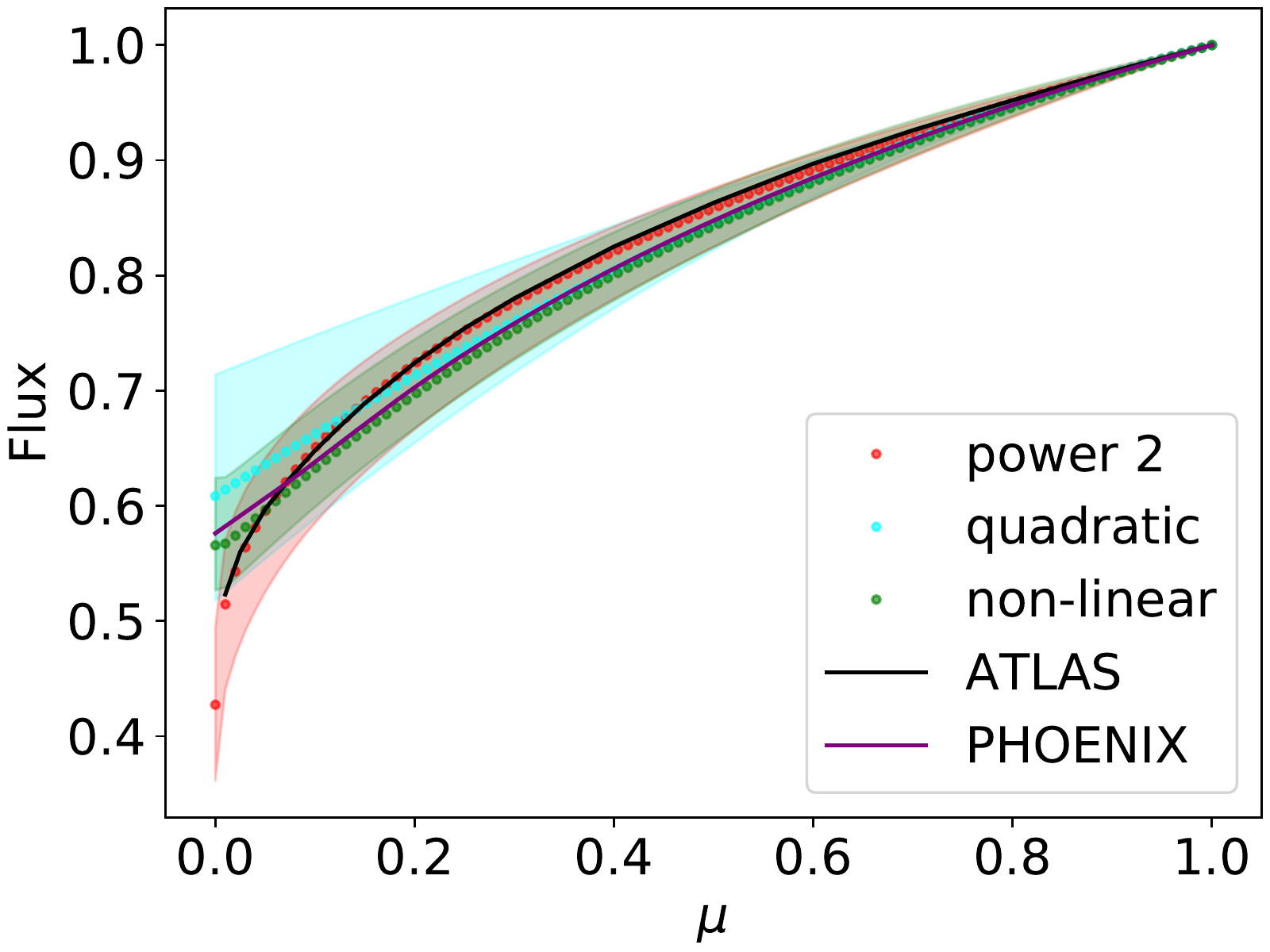}
\caption{Stellar intensity profiles from the PHOENIX (solid purple line) and ATLAS (solid black line) stellar grids for the HST WFC3.IR.G141 filter as a function of $\mu$ ($\mu =  \sqrt{1-z^2}$,  where z is the normalised distance from the centre of the stellar disc). We overlaid the best fit limb darkening models for the power-2 (dotted red), quadratic (dotted cyan), and non-linear (dotted green) laws. We also plotted the range of the parameter space allowed by the limb darkening models using the derived parameter uncertainties  after multiplying the intrinsic theoretical model uncertainties provided by LDTk by $40\times$ for the quadratic model and $10\times$ for the power-2 model. The intrinsic uncertainties of the modelled grids were not changed for the derivation of the non-linear LD parameters. \label{LDs}} 
\end{figure}

We assessed if the LD model affected the measurement of the Love number by performing several tests. We compared the results for three different LD laws: the quadratic law which is the most widely used, the non-linear 4-coefficient law considered to be the best model, and the power-2 law which has been shown to give good results despite its simplicity. We fitted the LD coefficients using priors derived from the stellar models. We found that the results depend on the priors. In particular, if the priors were very large, the results are independent from the stellar models. This results in a loss of information and loss of correlations between the four different colours which is not ideal since they relate to the same star. Therefore, to try to have LD coefficients that are consistent for our four instrument filters, we investigated which were the smallest reasonable priors for the parameters for each law. To achieve this, we compared the stellar intensity profiles from the ATLAS9 models \citep{Castelli2003} with the ones from the PHOENIX models \citep{Husser2013}. We used the LDTK code \citep{Parviainen2015} to fit the limb darkening laws mentioned above for the four filters increasing the intrinsic uncertainty of the models, which account for the uncertainty in the stellar parameters, in order to obtain modelled law uncertainties that encompass both the PHOENIX and the ATLAS stellar intensity profiles. This required increasing the intrinsic model uncertainties by $5-40\times$ for the quadratic and the power-2 law. The factor is higher for the visual filters than for the infrared. For the non-linear law, there is no need to increase the model uncertainties because the four parameters give sufficient flexibility for the model to encompass both sets of stellar intensity profiles. The uncertainty of the modelled LD law was derived by randomly drawing LD coefficients from a Gaussian distribution centred on the LD value and standard deviation equal to its uncertainty. An example, of the fit for the HST filter is shown in Figure~\ref{LDs}. For CHEOPS, we could not derive the intensity profile for the ATLAS models so we used the KEPLER filter, which is similar to CHEOPS, as a proxy for the uncertainties needed. Moreover, for Spitzer, we needed to redefine the stellar radius for the PHOENIX models in order to match the stellar radius of the ATLAS models since in this case the automatic limb definition does not give optimal results  \citep{Parviainen2015,Espinoza2015}.

From Figure~\ref{LDs}, it is clear that the PHOENIX and the ATLAS models predict different stellar intensity profiles close to the stellar limb. It is also clear that the power-2 law matches the ATLAS models better and that the non-linear law matches the PHOENIX models better (the latter is by construction). A comparison between LD derived from transit light curves and from theoretical models by \citet{Espinoza2015} suggested that the ATLAS models might be a better match to the transit fitted LD coefficients. However, this conclusion might depend on several factors that have yet to be investigated. Therefore, we expect that the quadratic law will be a poor description of the true stellar intensity profile and that the power-2 law will be a good description if the ATLAS models are closer to the true stellar intensity. We also expect that the non-linear LD law has enough flexibility to match both cases.

The uncertainties of the LD coefficients derived with the procedure described above were used to set the priors on LD coefficients for the transit light curve fit for the spherical and the ellipsoidal model. The quadratic LD coefficient priors are given in Table~\ref{prior_quadratic_ld}, while the non-linear LD coefficient priors are given in Table~\ref{prior_nl_ld}. The priors and results for the adopted model -- the power-2 law -- were already given in Section~\ref{deformation}. We find that despite its simplicity, the power-2 law gives results in good agreement with the more complex non-linear limb darkening law. For the three LD laws that we tested, we obtained consistent results with all of the fitted parameters agreeing within $1\, \sigma$. In Table~\ref{ldresutls}, we give the derived Love number, the significance of the detection, and the Bayes factors. As mentioned above, for model comparison, the Bayes factors should be multiplied by the prior odds that are very strongly in favour of the ellipsoidal model. The significance of the results varies slightly, but it agrees well between the models supporting the robustness of our results. For both the power-2 law and the non-linear law, we obtained a detection of the Love number of WASP-103b at more than $3\,\sigma$ and consistent with each other. It is noteworthy that although the quadratic law provides results compatible within $1\, \sigma$ with the other laws, it yields the smallest value and the largest uncertainties for $h_f$. This supports the idea that it is the worst model of the three. Since the three models agree well within $1\, \sigma$, we conclude that our treatment of the limb darkening is robust and it is not biasing the results.

If we increase the uncertainties of all the priors of the LD coefficients, for example to 0.1, we still obtain consistent values for $h_f$, despite, as expected, the detection significance being reduced to $\sim 2\, \sigma$. However, we think this overestimates the true uncertainty of the LD, especially in the infrared where the LD signature is much smaller. If we use the intrinsic uncertainties derived from the theoretical stellar grids that are much lower than the ones we derived with our method (up to $40\times$), we find Love number values that are too high, indicating that the LDs were biasing the retrieval of the Love number. Therefore, we find that the best approach is to use as much prior information from the theoretical stellar grids as possible, while taking the differences associated with different models into account (in our case ATLAS and PHOENIX).

\begin{table} 
\caption{Comparison of the derived Love number, significance of the detection, and Bayes factors for the three LD laws considered.}
\begin{tabular}{lccc} 
\hline 
\hline LD law & Love number & Significance & Bayes factor \\ 
Power-2 law &  $1.59^{+0.45}_{-0.53}$ & 3$\sigma$& $9.1(17^*)$ \\
Quadratic & $1.37^{+0.51}_{-0.59}$& 2.3$\sigma$ &4.6$(6.6^*)$ \\
Non-linear & $1.69^{+0.42}_{-0.48}$ & 3.5 $\sigma$& 16$(26.9^*)$ \\
\hline 
\hline 
\end{tabular} 
\tablefoot{\\
$^*$ Corrected value of the Bayes factor as explained in Section~\ref{signi}.}
\end{table}

\subsection{Future prospects of measuring the tidal deformation}

Since the Love number is an important constraint for interior models, we tested the possibilities of constraining it better. A higher significance of the result would also be desirable for a more robust detection which requires more high signal-to-noise transits of WASP-103b.
We simulated more seasons of CHEOPS observations, assuming in each season we would observe 12 more transits. If we could obtain four more seasons of observations ($48$ transits over 4 years), we would be able to measure the Love number of WASP-103b at $4.3\,\sigma$. If we could obtain six more seasons of observations (72 transits over 6 years), we would be able to derive  $h_f $ at $5\,\sigma$. This would require the CHEOPS mission being extended.

The extreme high precision of JWST and the fact that the limb darkening signature is lower in the infrared implies that the best chances of significantly increasing the precision of the measurement of the Love number in the near future is to combine our data with a transit with JWST. Since we are interested in maximising the cadence and the signal-to-noise of the observations, the best would be to use the NIRSpec Prism mode which would enable a precision of $62\,$ppm/min. We simulated a transit of WASP-103b with NIRSpec Prism mode assuming the Love number derived above.  We assumed a limb darkening profile similar to the one of Spitzer channel 1 since it has a similar wavelength range as the NIRSpec Prism. This simulated JWST transit was combined with our data and we followed the same procedure as above to derive the transit parameters. We obtained a $12\,\sigma$ detection of the Love number of WASP-103b, $h_f  = 1.62 _{-0.13}^{+0.12}$. This would be an unprecedented constraint on the Love number of an exoplanet  which would give us strong insights into the interior of these planets and their similarities and differences with the Solar System giants.

\subsection{Measuring the Love number from planet-planet interaction}
\label{precession}

HAT-P-13~b is the only exoplanet for which the measurement of the Love number was confirmed. For HAT-P-13~b, the determination of the Love number was made by an alternative method through the fixed point orbital eccentricity \citep{Batygin2009, Kramm2012, Buhler2016}. This method, proposed by \citet{Batygin2009}, is based on dynamical effects, and thus accesses the potential Love number, $k_f$, instead of the radial Love number, $h_f$, as in our case. The two Love numbers are related to each other by $h_f = 1 + k_f$  \citep[e.g.][]{Lambeck1980}, but while $h_f$ solely depends on the shape of the planet, $k_f$ depends on the knowledge of all dynamical effects in the system that can disturb the precession rate. It is, therefore, a much less direct method. 

 Measuring the $h_f$ directly from the signature in the light curve only assumes that the orbit of the planet is circular and its rotation synchronous \citep{Correia2014}. These two hypotheses are very likely for planets near the Roche limit. In contrast, many more assumptions are required to measure $k_f$ because the precession rate depends on many physical effects, namely general relativity, rotational flattening of both the star and the planet, tidal deformation of both the star and planet, and secular perturbations from all the remaining planets in the system. 
 
  \citet{Batygin2009} applied their method to the HAT-P-13 system, which is the only system currently known that fulfils the criteria of applicability. However, some assumptions were necessary for the derivation of the Love number. The most important are coplanar orbits, the eccentricity is locked in a fixed point, and a hierarchical two-planet system. These assumptions are consistent with current observations of the HAT-P-13 system, although they cannot be completely confirmed. If any one of the assumption fails, the measurement of the Love number of HAT-P-13~b would be biased. Nevertheless, the estimate of the Love number has allowed \citet{Batygin2009}, \citet{Kramm2012}, and \citet{Buhler2016} to place unprecedented constrains on the core mass and on the metalliticty of the planet's envelop, showing the potential of the Love number to lift degenerancies of the interior structure models. \citet{Csizmadia2019} applied the same method to WASP-18Ab deriving a Love number of $k_f = 0.62_{-0.19}^{+0.55}$. However, in this case, the cause of the orbital precession is not clear.

In conclusion, the apsidal precession method allows one to derive precise values for the potential fluid Love number for two planet systems with special orbital configuration. However, the several assumptions of the model can  have an impact on the accuracy of the measurements of the Love number.

\section{Conclusions}

We obtained 12 new high-precision transit observations of WASP-103b with the CHEOPS satellite to study the tidal interaction with its host star. The CHEOPS data were analysed with a multi-dimensional GP constrained by several instrumental parameters to correct the systematic effects due to the rotation of the field. We find that the roll angle, which measures the rotation of the field, is the instrumental parameter with higher correlation with the systematic effects. We also found that detrending the data with only the roll angle gives a good correction of the systematic noise. However, in most cases, including other instrumental parameters is a better model of the systematic noise according to Bayesian model comparison.

We find that a linear ephemeris is the preferred model for the orbital evolution of WASP-103b. However, there is a hint of an orbital period increase, contrary to what was expected if tidal decay was dominating the orbital period evolution of this planet. We explored scenarios that could explain a positive period derivative in case it is confirmed by future observations. 

One possibility is RV acceleration due to a bound companion. If the known visual companion of WASP-103 is bound, it could affect the transit times of WASP-103b. To check this, we obtained further RV observations with CORALIE and lucky imaging observations with the AstroLux camera. The RV observations are compatible with both a bound companion and a non-bound companion. We find an RV offset of $14 \pm 45$ m/s between the previous observations and the new observations spanning 8 years. This measured RV offset includes an unknown instrumental offset of 14-24 m/s and the hypothetical contribution from a bound companion. The value of the RV offset does not exclude that the visual companion is bound since the RV variations over the 8 year timescale of the observations are expected to be less than $23.2 \pm 3.3$ m/s. Although the RV required to cause the observed transit timing variations by the change in the light travel time is much higher ($342 \pm 146$ m/s), its high uncertainty also does not allow for the exclusion of this possibility. 

The new lucky imaging observations do not find the stellar companion despite the high sensitivity at the position it was observed before. To avoid detection, the companion star had to move in the direction of WASP-103 by 77 au, which is too large for a bound object. Therefore, the new lucky imaging observations support the idea that the visual companion is not bound unless unknown systematics have affected our results. Hence, our data support a non-bound companion, but we recommend further observations to confirm these results. Long-term monitoring of the RVs, as well as the new data release from \textit{Gaia}, can help constrain the visual companion. Furthermore, high resolution imaging  would allow confirmation of the position of the visual companion and its unbound nature.

Other possibilities to explain a positive period derivative are the Applegate effect and apsidial precession. However, a simpler explanation would be statistical artefacts. Several systematic effects have been shown to affect the measurement from transit times in exoplanets \citep[e.g.][]{Barros2013, Barros2020}. Hence,  statistical artefacts could cause the measured period to appear to be increasing. This is supported by the Bayesian evidence and a decrease in the measured period derivative $  \dot{P}  =  3.6 \pm 1.6 \times 10^{-10} $ relative to previous observations (  $  \dot{P}  =  8.4 \pm 4.0 \times 10^{-10} $  -- \citealt{Patra2020}). If we assume tidal decay is dominating the period evolution of WASP-103b, we can place a lower limit on the tidal quality factor $ Q'_* $  of $3.3 \times 10^6$ at $3\, \sigma$, corresponding to a 99.7\% confidence interval. This is similar to the recent limit on $ Q'_* $ obtained for WASP-18b $3.9 \times 10^6$ at a 95\% confidence interval \citep{Maciejewski2020}. For these systems, longer monitoring of the transit times will be required in order to constrain the stellar tidal quality factor.

We combined our new 12 CHEOPS light curves with previous transit light curves obtained by HST and Spitzer to measure the deformation of WASP-103b via its Love number. We re-reduced the light curves obtained with HST and Spitzer to correct for systematic effects since corrected light curves were not available in the literature. We measured the tidal deformation of the planet directly from its distortion of the transit light curve. We estimated the Love number of WASP-103b to be $h_f =1.59^{+0.45}_{-0.53}$, which is the first $3\,\sigma$ detection of an exoplanet Love number measured directly from its deformed transit shape.

The Love number of WASP-103b is similar to Jupiter's and slightly larger than Saturn's ($h_f =1.39 \pm 0.024$ -- \citealt{Lainey2017}). For a given planetary mass and radius, higher Love numbers imply a metal enrichment of the envelope and hence a decrease in the core mass. Our measurement of the Love number can be used to remove degeneracies in planetary internal models, allowing one to calculate the core size and the composition of WASP-103b \citep{Baumeister2020}. Uncertainties  in the limb darkening can influence the measurement of the Love number and hence we have performed a careful treatment of the limb darkening and several tests that indicate that our results are robust. 

Future observations with the JWST can help to better constrain the Love number of WASP-103b and gain an unprecedented view of the interior of this hot Jupiter. This could help us to better understand these extreme systems.

\begin{acknowledgements}
CHEOPS is an ESA mission in partnership with Switzerland with important contributions to the payload and the ground segment from Austria, Belgium, France, Germany, Hungary, Italy, Portugal, Spain, Sweden, and the United Kingdom. The CHEOPS Consortium would like to gratefully acknowledge the support received by all the agencies, offices, universities, and industries involved. Their flexibility and willingness to explore new approaches were essential to the success of this mission.
This work was supported by FCT - Fundação para a Ciência e a Tecnologia through national funds and by FEDER through COMPETE2020 - Programa Operacional Competitividade e Internacionalizacão by these grants: UID/FIS/04434/2019, UIDB/04434/2020, UIDP/04434/2020, PTDC/FIS-AST/32113/2017 \& POCI-01-0145-FEDER- 032113, PTDC/FIS-AST/28953/2017 \& POCI-01-0145-FEDER-028953, PTDC/FIS-AST/28987/2017 \& POCI-01-0145-FEDER-028987,UIDB/04564/2020
UIDP/04564/2020,  PTDC/FIS-AST/7002/2020, POCI-01-0145-FEDER-022217, POCI-01-0145-FEDER-029932. O.D.S.D. is supported in the form of work contract (DL 57/2016/CP1364/CT0004) funded by national funds through FCT. S.G.S. acknowledge support from FCT through FCT contract nr. CEECIND/00826/2018 and POPH/FSE (EC). 
MJH and YA acknowledge the support of the Swiss National Fund under grant 200020\_172746. 
SH gratefully acknowledges CNES funding through the grant 837319. 
D.K. acknowledges partial financial support from the Center for Space and Habitability (CSH), the PlanetS National Center of Competence in Research (NCCR), and the Swiss National Science Foundation and the Swiss-based MERAC Foundation. 
ACC and TGW acknowledge support from STFC consolidated grant number ST/M001296/1. 
PM acknowledges support from STFC research grant number ST/M001040/1. 
This work was also partially supported by a grant from the Simons Foundation (PI Queloz, grant number 327127). 
B.-O.D. acknowledges support from the Swiss National Science Foundation (PP00P2-190080). ABr was supported by the SNSA. 
We acknowledge support from the Spanish Ministry of Science and Innovation and the European Regional Development Fund through grants ESP2016-80435-C2-1-R, ESP2016-80435-C2-2-R, PGC2018-098153-B-C33, PGC2018-098153-B-C31, ESP2017-87676-C5-1-R, MDM-2017-0737 Unidad de Excelencia Maria de Maeztu-Centro de Astrobiologí­a (INTA-CSIC), as well as the support of the Generalitat de Catalunya/CERCA programme. The MOC activities have been supported by the ESA contract No. 4000124370. 
XB, SC, DG, MF, and JL acknowledge their roles as ESA-appointed CHEOPS science team members. 
This project was supported by the CNES. 
The Belgian participation to CHEOPS has been supported by the Belgian Federal Science Policy Office (BELSPO) in the framework of the PRODEX programme, and by the University of Liège through an ARC grant for Concerted Research Actions financed by the Wallonia-Brussels Federation. 
This project has received funding from the European Research Council (ERC) under the European Union’s Horizon 2020 research and innovation programme (project {\sc Four Aces} grant agreement No 724427). 
CMP and MF gratefully acknowledge the support of the Swedish National Space Agency (DNR 65/19, 174/18). 
DG gratefully acknowledges financial support from the Cassa di Risparmio di Torino (CRT) foundation under Grant No. 2018.2323 "Gaseous or rocky? Unveiling the nature of small worlds".
M.G. is an F.R.S.-FNRS Senior Research Associate. 
KGI is the ESA CHEOPS Project Scientist and is responsible for the ESA CHEOPS Guest Observers Programme. She does not participate in, or contribute to, the definition of the Guaranteed Time Programme of the CHEOPS mission through which observations described in this paper have been taken, nor to any aspect of target selection for the programme. 
Acknowledges support from the Spanish Ministry of Science and Innovation and the European Regional Development Fund through grant PGC2018-098153-B- C33, as well as the support of the Generalitat de Catalunya/CERCA programme. 
This project has been supported by the Hungarian National Research, Development and Innovation Office (NKFIH) grants GINOP-2.3.2-15-2016-00003, K-119517, K-125015, and the City of Szombathely under Agreement No.\ 67.177-21/2016. 
V.V.G. is an F.R.S-FNRS Research Associate. 
J.L-B. acknowledges financial support received from ”la Caixa” Foundation (ID 100010434) and from the European Union’s Horizon 2020 research and innovation programme under the Marie Skłodowska-Curie grant agreement No 847648, with fellowship code LCF/BQ/PI20/11760023.
Based on observations collected at Centro Astronómico Hispano en Andalucía (CAHA) at Calar Alto, operated jointly by Instituto de Astrofísica de Andalucía (CSIC) and Junta de Andalucía. GB acknowledges support from CHEOPS ASI-INAF agreement n. 2019-29-HH.0. ML acknowledges support from the Swiss National Science Foundation under Grant No. PCEFP2 194576. A.De. acknowledges the financial support of the European Research Council (ERC) under the European Union's Horizon 2020 research and innovation programme (project {\sc Four Aces}; grant agreement No 724427). A.De. also acknowledges financial support of the Swiss National Science Foundation (SNSF) through the National Centre for Competence in Research "PlanetS". L.D. is an F.R.S.-FNRS Postdoctoral Researcher. We thank Daniel Foreman-Mackey for his insight into model comparison versus parameter inference. 
\end{acknowledgements}

\bibliographystyle{aa} 
\bibliography{susana}

\appendix

\section{Figures of the lucky imaging observations}

\begin{figure}[ht!]
\centering 
\includegraphics[width=0.9\columnwidth]{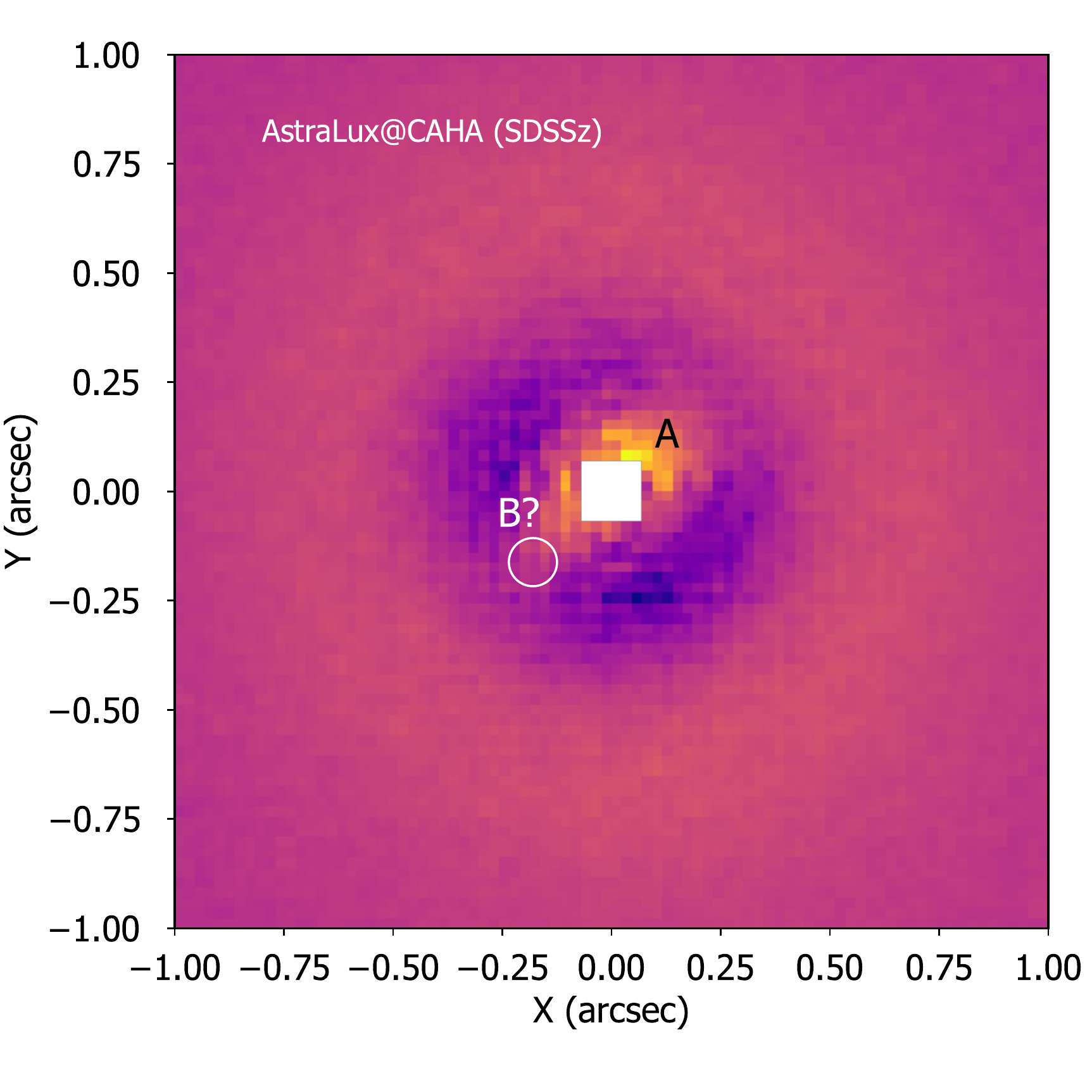}
\caption{ AstraLux high-spatial resolution image of WASP-103 obtained on 13 January 2021 in the SDSSz bandpass. The image corresponds to the stacking of the 10\% frames with the highest Strehl ratio. We removed a fitted PSF of the main target as a combination of a Gaussian and a Lorentzian profile. The location of the previously detected companion by \citet{Ngo2016} is marked as 'B?' .\label{HRimage} } 
\end{figure}

\begin{figure}[ht!]
\centering 
\includegraphics[width=0.9\columnwidth]{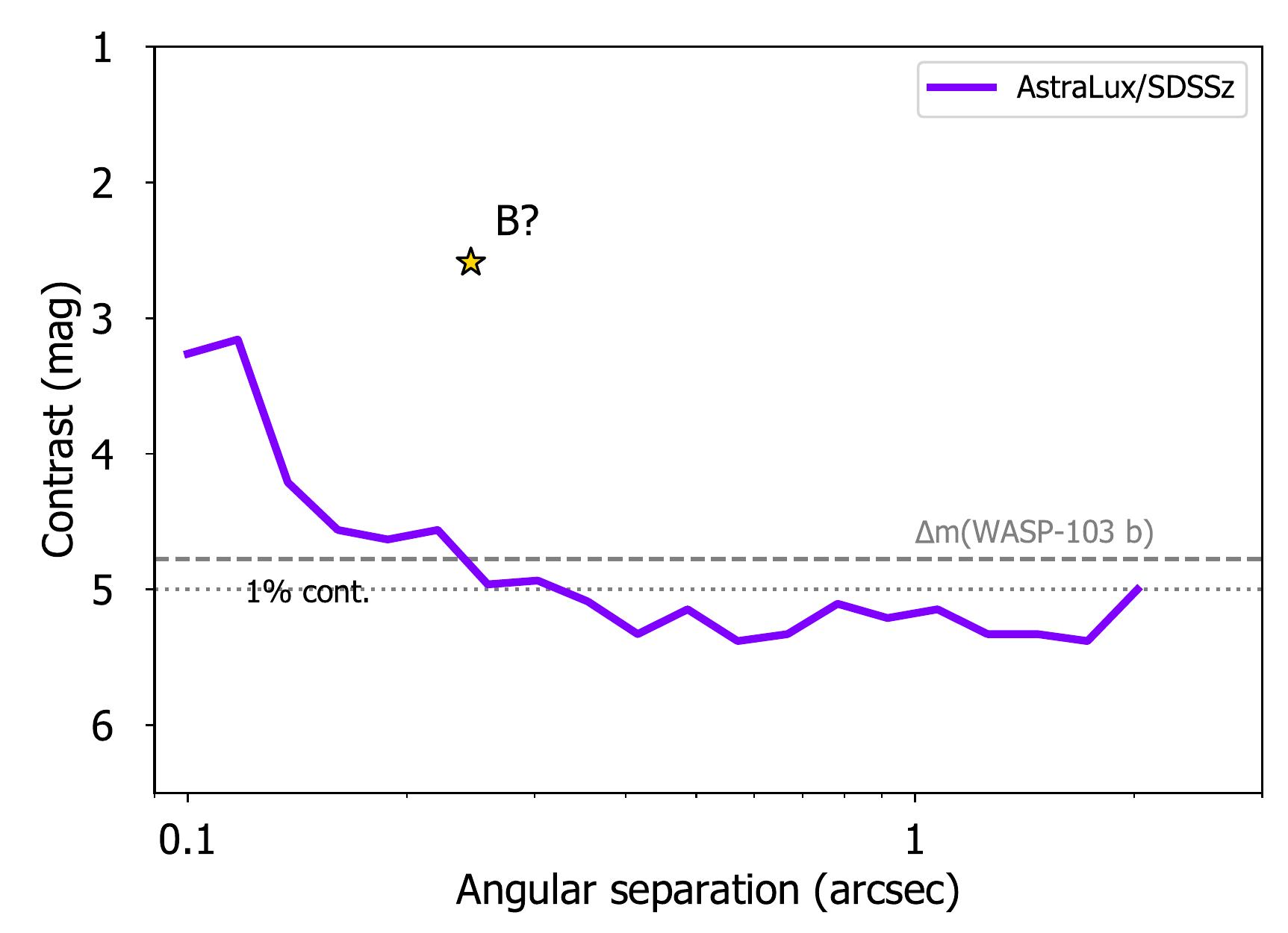}
\caption{ Sensitivity curve for the AstraLux image of WASP-103. The 1\% contamination level is marked by the horizontal dotted line and the maximum magnitude of a blended binary to be able to mimic the transit of WASP-103b is marked by the horizontal dashed line. The location of the previously detected companion by \citet{Ngo2016} is marked as a star-like symbol.  \label{Sensitivity} } 
\end{figure}

\section{Priors for the limb darkening coefficients}

\begin{table}[ht!]
\caption{Priors for LD coefficients for the quadratic law. \label{prior_quadratic_ld} }
\begin{tabular}{lc} 
\hline 
\hline Parameter & Prior  \\ 
$LD1$ CHEOPS&  $\mathcal{N}( 0.5269, 0.0218)$  \\ 
$LD2$ CHEOPS &  $\mathcal{N}( 0.1279, 0.046)$  \\ 
$LD1$ HST&  $\mathcal{N}(0.2346, 0.0074)$   \\ 
$LD2$  HST &  $\mathcal{N}(  0.1461, 0.0266)$  \\ 
$LD1$ SP1&  $\mathcal{N}(0.1080, 0.018)$   \\ 
$LD2$  SP1 &  $\mathcal{N}(  0.1220, 0.0268)$  \\ 
$LD1$ SP2&  $\mathcal{N}( 0.0920, 0.0138)$   \\ 
$LD2$  SP2 &  $\mathcal{N}(0.0976, 0.0214)$  \\ 
\hline 
\hline 
\end{tabular} 
\tablefoot{\\
$\mathcal{N}(a;b)$ is a normal distribution with mean $a$ and standard deviation $b$. }
\end{table}

\begin{table}[ht!]
\caption{Priors for LD coefficients for the non-linear law. \label{prior_nl_ld} }
\begin{tabular}{lc} 
\hline 
\hline Parameter & Prior  \\ 
u CHEOPS&  $\mathcal{N}( -0.0863 ,0.0089)$  \\ 
v CHEOPS&  $\mathcal{N}( 0.8962 ,0.0147)$  \\ 
w CHEOPS&  $\mathcal{N}( -0.0438 ,0.0149)$  \\ 
z CHEOPS&  $\mathcal{N}(-0.1310 ,0.0065)$  \\ 
u HST&  $\mathcal{N}(-0.1113  ,0.0084)$  \\ 
v HST&  $\mathcal{N}(1.3060  ,0.0113)$  \\ 
w HST&  $\mathcal{N}(-1.0551  ,0.0076)$  \\ 
z HST&  $\mathcal{N}( 0.2854  ,0.0026)$  \\ 
u SP1&  $\mathcal{N}( -0.0074   ,0.00258)$  \\ 
v SP1&  $\mathcal{N}(0.7294   , 0.00412)$  \\ 
w SP1&  $\mathcal{N}( -0.7218  ,0.00332)$  \\ 
z SP1&  $\mathcal{N}(0.2426   ,0.0017)$  \\ 
u SP2&  $\mathcal{N}( 0.0078   ,0.00234)$  \\ 
v SP2&  $\mathcal{N}(  0.5679   , 0.00418)$  \\ 
w SP2&  $\mathcal{N}(  -0.5661   ,0.00362)$  \\ 
z SP2&  $\mathcal{N}(  0.1932   ,0.00162)$  \\ 
\hline 
\hline 
\end{tabular} \\ 
\tablefoot{\\
$\mathcal{N}(a;b)$ is a normal distribution with mean $a$ and standard deviation $b$.  }
\end{table}

\end{document}